\documentclass{aa}  

\usepackage{epsfig}
\usepackage{amsmath}
\usepackage{amssymb}
\usepackage{graphicx}
\usepackage{txfonts} 
\usepackage{natbib}
\bibpunct{(}{)}{;}{a}{}{,} 
\usepackage{hyperref}
\usepackage[flushleft]{threeparttable}
\usepackage[normalem]{ulem}
\usepackage{xcolor}
\usepackage{siunitx}

\defcitealias{lynden-bell72}{LBK72}
\defcitealias{meidt24}{MvdW24}
\defcitealias{van-der-wel25}{vdWM25}

\begin{document} 

   \title{Nexae in caverna: the secular evolution of disks via collectively excited, transient spiral structure}
   \titlerunning{Nexae in caverna}

   \author{Sharon E. Meidt\inst{1}
          \and
          Arjen van der Wel\inst{1}
          }

   \institute{Department of Physics and Astronomy, Universiteit Gent, Krijgslaan 281 S9, B-9000 Gent, Belgium\\
              \email{sharon.vanderwel@ugent.be}
             }

   \date{Accepted 17 August 2026}
 
  \abstract{}{
Using the hydrodynamical (fluid) approximation, we present a self-consistent theoretical framework that couples the origin, evolution, and decay of spiral structures to the secular dynamical evolution of their host galactic disks. }
{Our approach highlights non-resonant spiral excitation through azimuthal forcing that leverages mild, pervasive gradients in the disk's mass and angular momentum distributions, structural features we term \textit{cavernae}. These cavernae are weaker but more widespread than the sharp features behind traditional groove mode excitation and commonplace in exponential disks. We discuss how non-resonant features combine with the other responses in exponential disks---resonant dressing, steady waves, groove modes---to produce a global, evolving spiral \textit{nexum} that transports angular momentum and reshapes the disk. 
  }
  {Using expressions for spiral torques, angular momentum transport and heating, we demonstrate that global spirals are intrinsically self-limiting; the angular momentum changes and heating they generate rapidly quenches their own growth, dictating a finite lifetime for any single spiral episode. A succession of transient episodes, each with properties adjusted to the changed disk conditions, lays the pathway to long-lived spiral activity. This behavior suggests that the character of secular evolution shifts over time. We find that the short-lived, high-multiplicity (high-$m$) spirals that dominate in dynamically cold disks induce widespread, impulse-like non-resonant heating, yet with a low ratio of heating to radial migration. As the disk warms, these high-$m$ features are suppressed, leading to steadier, lower-$m$ spirals that heat the disk progressively more efficiently near discrete resonances.} {In this light, the dynamical coldness of many observed disk galaxies today requires a past dominated by high-$m$ transient perturbations, whereas warmer, more compact systems reflect an advanced stage of secular evolution regulated by transient, low-$m$ spirals. Ultimately, the heating and reshaping produces stable (featureless), centrally concentrated, but disky/rotating early-type galaxies.}


   \maketitle
   
\section{Introduction}
The origin and persistence of spiral structures in galactic disks remain a long-standing puzzle in extragalactic astronomy \citep[for reviews, see][]{dobbs14, sellwood14a, sellwood22a}. Their prominence across nearly the entire electromagnetic spectrum \citep[e.g.][]{elmegreen84, kennicutt03} gives these features an undeniable aesthetic appeal that has inspired decades of observational, numerical, and analytical study. However, the precise mechanisms by which these patterns originate, evolve over secular timescales, and eventually fade remain subjects of ongoing investigation \citep[e.g.,][]{sellwood11, sellwood12, meidt24, hamilton24}.

Underlying this puzzle is a fundamental theoretical divide over whether spiral arms are enduring structures or temporary, shearing patterns. On one side is the preference for long-lived density wave modes, rooted in the Quasi-Stationary Spiral Structure (QSSS) hypothesis and later global modal theories \citep[e.g.][]{lin64, bertin89, zhang96,zhang98,zhang99}. On the other side is the argument for transient, shearing material patterns \citep{julian66,toomre81,baba13}, which take an arguably more direct route to prominence via swing amplification \citep{goldreich65, toomre91}.

Despite the elegance and efficiency of swing amplification, steady, long-lived waves have maintained a strong conceptual foothold in the literature. Much of this enduring appeal can be traced back to the seminal calculations of \citet[][hereafter, LBK72]{lynden-bell72}. Using an action-angle formalism, LBK72 demonstrated how steady spirals drive the secular rearrangement of angular momentum within their host galaxies at specific resonances. Those calculations anchored the preference for quasi-stable disks, forming the theoretical basis for our modern understanding of how spirals shape galaxy evolution over their lifetimes.

Revisions to this classical picture have emerged more recently, driven largely by insights from numerical simulations \citep{sellwood11,fujii11,sellwood14, sellwood19, roskar12, grand13, baba13, grand26}, which consistently demonstrate that spirals behave as transient, recurrent features. These simulations are increasingly paired with dedicated analytical calculations across specific regimes—-spanning short and long waves, transient and steady spirals, and interactions near corotation or the Lindblad resonances \citep[e.g.,][]{barbanis67, binney88, jenkins90, sellwood91,sellwood02, hamilton24}.  Together these establish how the heating and angular momentum exchanges driven by a given spiral dictate its own decay, while simultaneously shaping the disk's susceptibility to future generations of spirals—effectively framing a feedback loop between recurrent waves and the host disk.

Parallel to these efforts, another class of analytical models attempts a more holistic view of secular evolution. These approaches frame disk dynamics in terms of changes to the backbone of invariant orbital tori \citep{lau21a}, stochastic diffusion through action space \citep[e.g.,][]{fouvry15}, or the direct evolution of the distribution function \citep{hamilton24, hamilton25}. While theoretically robust, these methods can sometimes obscure the physical link between the large-scale morphological transitions of the disk and the observable properties of the spirals as they grow and decay. Consequently, critical questions remain open regarding the true lifetimes of spiral patterns and the extent to which they dynamically reshape their host galaxies.

For example, the stochastic diffusion framework \citep[e.g.,][]{fouvry15} elegantly captures the basic idea behind the disk-spiral feedback loop (called 'dressed diffusion'): spiral irregularities emerge as (resonant or non-resonant dressing), stars "scatter" off the spiral irregularities in action-space and this in turn stifles the instabillity itself.
However, the complexity of the Balescu-Lenard equations makes them computationally intensive to solve and few of the analytical calculations of diffusion completed to date apply specifically to disks (in which case the diffusion corresponds to radial migration and heating) or proceed outside the tightly-wound limit \citep[see][]{lynden-bell72, fouvry15, heyvaerts17, roule25}.
Although these calculations imply an important role for short-lived spirals in disk heating, swing-amplification and non-resonantly growing collectively excited open dressed modes are difficult to characterize in this context.
Approaches like this tend to miss the full impact of short-lived (transient) spirals \citep{van-der-wel25} as they must contend with the fact that self-consistent solution to the stellar dispersion relation is difficult and analytically complicated for damping modes (although approaches have been suggested for treating weakly damped modes \citep{weinberg04}).

The galactokinetics approach by \citet{hamilton26} is designed to provide a much more intuitive, broadly influential view, capturing the decay of spirals with a given set of properties as a transport process in action space.
This makes it extraordinarily valuable for describing now common-place signatures of phase mixing in the MW (e.g. \citealt[][]{hamilton26a}; see also \citealt{antoja18,antoja22,frankel25,alinder23}).
On the other hand, it can be 
difficult to map this approach's microscopic view of spiral evolution onto the properties of the spirals themselves or to the macroscopic changes they produce in their host galaxies.

The hydrodynamical approach we have been employing in a series of papers bypasses many of these issues \citep[][hereafter, MvdW24 and vdWM25, respectively]{meidt24, van-der-wel25}.
Although historically analytical hydrodynamical calculations have played an integral part in building our basic picture of spiral evolution, they have recently fallen out of favor compared to stellar dynamical calculations. Yet, direct access to the time evolution of the fluid density  -- as opposed to the distribution function -- circumvents complications with damping in stellar treatments, allowing for a description of growth and decay of perturbations.
\footnote{With a hydrodynamical approach, the time evolution of the 
perturbed density and potential are easily accessed by combining the 
continuity equation with Poisson's equation. In the stellar dynamical 
approach, an equation for the evolution of the distribution function 
needs to be Laplace transformed to obtain the density and potential, 
severely limiting access to damping solutions. Although the hydro 
approach loses its accuracy below the Jeans length and 
inherently smooths over fine-grained, microscopic kinetic phenomena
such as resonant orbit trapping and stellar phase-space 
mixing, it yields powerful insights for 
the global behavior of large-scale spiral features.}

By treating the stellar disk as a fluid with an effective pressure \citep[e.g.,][]{lin64,goldreich65}, we replace the formal density-space complexities of kinetic theory with a single macroscopic variable: velocity dispersion (modeled as fluid pressure).
This makes it easier to track the transport of angular momentum and the resulting "heating" of the disk.
Although the microphysics below the Jeans length is no longer directly accessible, the hydrodynamical approximation provides a highly desirable macrophysical view of the feedback loop between spirals and galaxy disks.

The goal of this work is to use our hydrodynamical framework to obtain a transparent view of the secular disk changes produced by spirals, showing how and where they lead to transience.
We will highlight the manner in which the fluid approximation retrieves several classic results from stellar calculations and numerical simulations.
We will also relate non-resonant spiral amplification to mild gradients in potential vorticity (borrowing the term from \citet{papaloizou89} and studies of Rossby wave instability, e.g. \citealt{lovelace78, li00}), or guiding center distribution (in the context of stellar disks), which can be influential away from resonance for non-steady waves.
In this light, we show that exponential disks readily naturally harbor widespread deficits in their angular momentum distributions -- a feature we term \textit{cavernae} -- which are broader, milder cousins of sharp grooves in Mestel disks \citep{sellwood89}. Cavernae produce non-resonantly, collectively growing spirals by leveraging "donkey" behaviour in the orbiting stars (which we term \textit{nexae caverna}). 
In earlier hydrodynamical calculations, the mild vorticity gradient was considered for its influence either only at corotation \citep{mark76, goldreich78, goldreich79} or as an ingredient that produces steady spiral modes \citep{bertin88, bertin89, bertin89a}.

The calculation we use applies equally well to perturbed stellar disks (above the Jeans length, where the hydro approximation is valid) as it does to perturbed gas disks.
In a forthcoming paper we will make heating in gas disks, combined with dissipation, the principal focus.
The primary aim is to build a foundation for treating the rich multi-scale gas structure witnessed by MIRI \citep{meidt23} as a source of turbulent motion, and combine it with dissipation to predict an equilibrium velocity dispersion in gas disks.

To capture the full evolutionary arc of these structures, this paper is organized sequentially. In Section \ref{sec:growth_rates}, we establish the characteristic equation for spiral amplification, define the architecture of the caverna, and map our fluid conditions to classic stellar dynamical instabilities. In Section \ref{sec:timedep}, we isolate the properties of the fastest-growing waves to build a heuristic, time-dependent model of a single spiral episode from birth to peak saturation. In Section \ref{sec:heating}, we translate this peak potential into macroscopic equations for time-integrated torques and widespread non-resonant dynamical heating. Finally, in Section \ref{sec:disk_evolution}, we calculate the lifetimes of these spirals, demonstrating how the heat they generate dictates their own decay, driving a long-term spectral coarsening that transforms flocculent, dynamically cool disks into quasi-steady, grand-design morphologies.

\section{The elements of dynamical heating}\label{sec:elements}

\subsection{Treatments of heating in stellar disks}
The shaping of the morphology and evolution of stellar disks by collectively excited spiral structures is a longstanding area of study that has been approached from both stellar dynamical \citep[e.g.,][]{kalnajs71, toomre81, binney88,jenkins90,sellwood02, binney08} and numerical \citep[e.g.,][]{roskar08,sellwood11,fujii11,roskar12, solway12,sellwood14} perspectives.
The approach here seeks to re-envision those basic ideas from the hydrodynamical perspective, to make the decay of spirals more directly related to macroscopic, easily observable, disk quantities.
Thus, instead of describing the disappearance of spirals as a reshaping to increasingly small scales by shear, phase mixing is recast as changes in angular momentum and an increase in velocity dispersion (pressure) that then overcomes self-gravity to lead to the decay of the spiral.

The advantage of such a hydrodynamical view is that it provides a bridge between stellar dynamical calculations and numerical simulations and offers more flexibility in terms of the diversity of host disk conditions that can be considered.
We will show later in Section \ref{sec:amp_systems}, for example, the differences in spiral growth and evolution in Mestel disks compared to exponential disks.
More influentually, a hydrodynamical treatment also makes it easy to recognize and treat non-adiabatic spirals that are able to grow and decay away from corotation.
That growth was discussed by \citet{van-der-wel25}, who argued that such non-adiabatic features are missed by secular calculations sensitive to the behavior over longer timescales \citep[see also][]{banik21}.

Here we examine the implications of non-adiabatic growth (the `non-resonant dressing'), which necessarily exerts torques and heats the disk and thus necessarily leads to transience.  Combined with angular momentum changes produced by structures centered around corotation (the 'resonant dressing'; \cite{sellwood02,sridhar19,hamilton24}), the result is that spiral density waves are short-lived phenomena (see \cite{sellwood11,sellwood12,sellwood14}), as opposed to the quasi-steady structures envisioned by \cite{lin64}, \cite{bertin88} or \citet{zhang98}. 
At the same time, the heating and angular momentum changes primes the disk for new generations of spirals, ultimately giving rise to a wide hollow in the disk distribution function.
The hollow or 'caverna' is the cousin of grooves in Mestel disks found to give rise to spirals in numerical simulations \citep{sellwood89, sellwood91, sellwood12, fouvry15, de-rijcke19} but found here to be a more general feature of disks, including exponential disks embedded in a dominant external potential.

In the following sections we will assemble the ingredients that support this view, using the characteristic wave equation to identify the conditions for spiral amplification in section \ref{sec:growth_rates} and then using the patterns of growth to calculate the rates of angular momentum changes and heating due to spirals in section \ref{sec:heating}. To set up those components, below we first summarize the perturbations.

\subsection{Growth and decay from a hydrodynamical perspective}\label{sec:hydro}
In conventional approaches to determining spiral (in)stability, the linearized equations of motion and the Poisson equation are transformed into an eigenvalue problem that is solved by projecting the system onto a set of bi-orthogonal basis functions \citep[e.g.][]{kalnajs77, papaloizou89, sellwood01,de-rijcke16}.
Although the perturbations in this approach are time-dependent, the background state is time-independent.
Given that the problem at hand revolves around a variable rather than a fixed background state, we wish to approach the stability of spiral waves from a more general perspective, rather than seeking the spectrum of modes to be recalculated at each time.
That is, we wish to keep a transparent view of the factors that determine the mode spectrum at any given time, rather than the structure of the modes themselves.

To this end, we employ a strategy with three main features.
First, we rely on a picture of stability and instability portrayed by a characteristic equation for the doppler shifted frequency $(\omega-m\Omega$) (MvdW24), rather than a picture of mode structure derived from solutions to a wave-like equation.
Second, rather than solve the Poisson integral to link the potential perturbation to the density, we seek the qualities needed by the potential-density relation in order to yield stability, growth or decay.
Third, we replace precise knowledge of the potential-density relation at all times with a generic model for the time-dependence of potential as a growing and decaying perturbation.
The remainder of this section lays the ground work for the these strategies.

\subsection{Potential perturbations}\label{sec:potential_perts}
To calculate the secular evolution produced by stellar spirals we treat each as a small quasi-periodic perturbation to the underlying axisymmetric disk.
The potential of each structure is written as the sum of quasi-periodic perturbations of the form 
\begin{equation}
\Phi_{1,m}(R,\phi,z,t)=\Re{[\mathcal{F}(R,z)e^{i(m\phi-\int\omega dt)}e^{ik R }]}  \label{eq:pertex}
\end{equation}
where $\omega=m\Omega_p+i\beta$ is the perturbation's complex frequency, and the spiral multiplicity $m$ and the radial wave number $k$ are related through $k= m\tan {i_p}/R$ where $i_p$ is the spiral pitch angle.

We adopt this form for the self-gravity perturbations and the external perturbations implicitly, as well (i.e. viewing any (compact) object in terms of a sum of independent Fourier components, as in the case of a Gaussian with a periodic extension, for example). As in \citetalias{meidt24,van-der-wel25}, our greatest interest is in the evolution of the self-gravity perturbation, and we will primarily focus on the period of time when self-gravity exceeds the external background potential (section \ref{sec:growth_rates}).  Prior to this time, the basic idea is that an external perturbation is introduced into the system and this stimulates an initial response in the disk.  External perturbation(s) could consist of giant molecular clouds, dark matter halo substructure, gas accretion events, as well as earlier spiral generations; e.g. \citealt{julian66,toomre72,fujii11,donghia,chakrabarti09,Kazantzidis08,Kazantzidis09}). The initial response to the external perturbation in many cases is a wake \citep[e.g.][]{goldreich65,julian66,toomre91}, but 
ultimately the conditions in the disk (rotation, pressure, self-gravity) determine the subset of Fourier components in the self-gravity spectrum that amplify to prominence  (i.e. \citetalias{meidt24, van-der-wel25}, and discussed again in section \ref{sec:growth_rates} using the characteristic equation derived from the linearized equations of motion).  At this stage, the form of the self-gravity perturbation no longer necessarily resembles that of the seed perturbation or initial wake.\footnote{
 In this light, 
the continued spiral pattern observable in the simulations of \citet{donghia} even after removal of the seed perturbation is a product of linear, self-gravity dominated evolution depicted by our approach.  The periodic nature of our calculation in cylindrical coordinates highlights a transition from local wake to $m$-periodic perturbation.  The recurrent nature of the spirals in the simulations of \citet{donghia}, or \cite{fujii11}, on the other hand, may be attributable to the non-linear terms in the continuity equation \citep[but see][]{sellwood21}.  This will be discussed more in section \ref{sec:nonlinearev} (see e.g. eq.[\ref{eq:s0sqlongNL}]).}
Changes to disk conditions brought about by the spiral itself in turn produce a quenching of its linear evolution, as discussed in sections \ref{sec:timedep}, \ref{sec:heating} and \ref{sec:disk_evolution} (see also \citealt{fujii11,sellwood21,hamilton24}).  

\subsubsection{The potential-density relation}
The relation between the potential and the density of the perturbations in eq. (\ref{eq:pertex}) is specified by Poisson's equation
\begin{equation}
\nabla^2\Phi=4\pi G\rho\label{eq:poisson}.
\end{equation}
The potential-density relation is an ingredient for predicting the growth and decay of perturbations from the (quasi) linear equations of motion and is essential for relating angular momentum exchanges and heating to the observable perturbation density (rather than potential).
Below we briefly outline considerations that determine the relation between density and potential in practice.
We will examine density perturbations, related to the potential perturbations $\Phi_1$, that are 3D, rather than restricted to the plane (2D), i.e. 
\begin{equation}
\rho_{1,m}(R,\phi,z,t)=\Re[\rho_{a,m}(R,z)e^{i(m\phi-\int\omega dt)}e^{ik R }],  \label{eq:pertdenseex}
\end{equation}
in which case the (separable) potential perturbation is written 
\begin{equation}
\Phi_{1,m}(R,z) = -\frac{2\pi G}{\vert k_t\vert} \int_{-\infty}^{\infty} \rho_{1,m}(R,z') e^{-|k_t||z - z'|} dz' \label{eq:poissonintegral}
\end{equation}
where $k_t$ is defined by the non-vertical terms in Poisson's equation, as discussed below \citep[also see][]{vandervoort70}.

\subsubsection{WKB and non-WKB perturbations}\label{sec:wkb}
In the most commonly adopted (and significantly simplifying) WKB approximation, the amplitude variation $d\mathcal{F}/dR$ is negligible.
In this case, we use eq. (\ref{eq:poissonintegral}) to find
\begin{equation}
\mathcal{F}(R,z)=-\frac{4\pi G\rho_{a,m}(R,0) z_0\mathcal{R}}{\vert k_t\vert }e^{-\vert k_t\vert z}
\end{equation}
where 
\begin{equation}
k_t=\sqrt{k^2+m^2/R^2}
\end{equation}
and in terms of the reduction factor  
\begin{eqnarray}
\mathcal{R}(k_t) &=& \int_{0}^{\infty} \textrm{sech}^2(z) e^{-\vert k_t\vert z} dz\nonumber\\
&\approx& \frac{1}{1+k_t z_0}
\end{eqnarray} 
calculated assuming a vertical density distribution
\begin{equation}
\rho_{a,m}(R,z)= \frac{\Sigma_{a,m}}{2z_0} \textrm{sech}^2\left(\frac{z}{z_0}\right).
\end{equation} 
Here and elsewhere we will assume that the spiral falls in the limit $k_t z_0\ll1$, in which case $\mathcal{R}\approx 1$.
Note, though, that as the vertical extent increases, a higher density would be needed to produce the same degree of response in the plane \citep[e.g][]{julian66,romeo92,ghosh18}.

For other, non-WKB perturbations the relation between the density and potential must reflect non-negligible gradients in the amplitude of the potential perturbation and its radial derivative, which evolve in time as the wave grows.
That is, we have
\begin{equation}\label{eq:t1def}
T_1 = \frac{d \ln \mathcal{F}(R,z)}{dR} =  \frac{d \ln \mathcal{F}_0(R,z)}{d R}+\int\frac{d \beta}{dR} dt,
\end{equation}
in terms of any gradient in the initial amplitude $F_0(R,z)$ and growth rate $\beta$, and 
\begin{equation}
T_2 = \frac{1}{\mathcal{F}} \frac{d^2 \mathcal{F}}{dR^2} = T_1^2 + \frac{d T_1}{dR}.
\end{equation} 

With these definitions, the general relation between the density and potential can be conveniently written as
\begin{equation}
\Phi_{1,m}(R,z)=-\frac{2\pi G\Sigma_{1,m}(R)}{\vert k_{t,e}\vert}e^{-\vert k_{e,R}\vert z}
\end{equation}
where $\Sigma_{1,m}=\rho_{1,m}(0) 2 z_0$ and
\begin{equation}
k_{\rm t,e}^2=k_{\rm e,R}^2+ik_{\rm e,I}^2
\end{equation}
with 
\begin{equation}
k_{\rm e,R}^2=\Re(k_{\rm t,e}^2)=k^2+\frac{m^2}{R^2}-\frac{T_1(R)}{R}-T_2(R)
\end{equation}
and 
\begin{equation}
k_{\rm e,I}^2=\Im(k_{\rm t,e}^2)=\left(\frac{k}{R}+2T_1k\right).
\end{equation}
Here again we assumed that $k$, $m/R$ and $T_1$ are all much smaller than $1/z_0$.

This complex quantity can be rewritten as a phase lag between the density and potential.
The behavior is well captured by 
\begin{equation}\label{eq:pot_dens_rel}
\Phi_{1,m}(0)=-\frac{2\pi G\Sigma_{1,m}}{k_{\rm e,long}}e^{-i\delta}
\end{equation}
where 
\begin{equation}
\frac{1}{k_{\rm e,long}}=\frac{\left(\Re{(k_{\rm t,e}^2)}^2+\Im{(k_{\rm t,e}^2)}^2\right)^{1/2}}{\vert\Re{(k_{\rm t,e}^2)}\vert^{3/2}+\frac{2}{3}\vert\Im{(k_{\rm t,e}^2)}\vert^{3/2}}\label{eq:kelong}
\end{equation}
and using a simpler formulation for the phase lag 
\begin{equation}
\delta=\arctan{\left(\frac{\Im{(k_{\rm t,e}^2)}}{\Re{(k_{\rm t,e}^2)}}\right)}.
\end{equation}

Throughout we assume both that $d\ln k/d\ln R$ and $d\ln T_1/d\ln R$ are negligible so that the wave form does not vary so rapidly across the disk that it is no longer a recognizably `quasi-uniform' structure.
In this case, $T_2=T_1^2+dT_1/dR\approx T_1^2$.

The phase lag $\delta$ between the density and potential introduced by the long-wave contributions to Poisson's equation allows open spirals to transport angular momentum \citep{zhang96, binney08, dehnen25}.
It is worth noting that the time-averaged torque density 
\begin{equation}
\langle \tau \rangle =-\int_0^{2\pi/m}\rho_{1,m}\frac{d \Phi_{1,m}}{d \phi}d\phi
\end{equation}
is negative inside corotation when the phase offset $\delta$ is negative or $k(1/R+2T_1)<0$.

\subsubsection{Approaches to approximate the potential-density relation outside the WKB limit }
The instantaneous rate of growth for any perturbation \citepalias[][see next section]{meidt24} depends on the potential-density relation and thus on $T_1$ (and $T_2$), which in turn evolve to reflect this growth.
To solve for the temporal evolution of any perturbation and the overall diffusion it produces \citep[e.g.,][]{fouvry15} thus requires self-consistent calculation of the evolution in both the disk and perturbations.

Complimentary but considerably simpler calculations of the torque and heating at any given moment in time \citep{binney08, dehnen25, van-der-wel25}, especially at peak spiral prominence, however, can provide valuable insight.
In such cases, the actual amplitude of a non-WKB potential perturbation and its radial variation in principle remains an important ingredient, given that the perturbed density is the observable quantity (as opposed to the potential).

A number of approaches have been devised and adopted, including next-order in the WKB approximation \citep{goldreich79, bertin89a} or the adoption of a particular form for the amplitude, like a power-law or an exponential \citep{dehnen25}.
The initial amplitude $F_0(r,z)$ could be assumed to vary no more rapidly than the unperturbed disk so as not to violate the assumption $\rho_1\ll\rho_0$ adopted in deriving the linearized equations of motion \citep{meidt22}.
Still other imagined perturbations include those that are finite in extent, in which case the perturbation's amplitude can be completely independent of the disk (without violating the assumption $\rho_1\gg\rho_0$).
These can satisfy the WKB approximation even for $kR\sim m\approx 1$.
As discussed in the next section, WKB-like perturbations are initially the fastest growing, making this a likely quality at early stages for spirals that grow to prominence.

Later in \S \ref{sec:timedep} we avoid the need to directly specify the potential's amplitude variation (or the potential-density relation) by invoking a model for the time evolution of spiral perturbations that links the initially fastest growing stage with the spiral at its peak.
The adopted model absorbs the unknown $T_1$, which implicitly determines how long a given spiral lives.
Before developing that model, we first discuss how we use the characteristic equation derived from the linearized equations of motion to identify the conditions associated with fastest and peak growth.

\section{Nexae in caverna: (Quasi-) Linear Growth rates and patterns of spirals in generic gas and stellar disks}\label{sec:growth_rates}
Perturbations of the form in eq. (\ref{eq:pertex}) are not only practical, they solve the linearized Euler equations of motion in a 3D perturbed differentially rotating disk.
In this section we will summarize the basic qualities of the characteristic equation derived with the linearized equations of motion (first introduced in \citetalias{meidt24} and again discussed in \citetalias{van-der-wel25}), and then describe what those qualities imply about spiral growth and decay.

We obtain a quasi-linear view from the picture of linear evolution by allowing that disk quantities may vary slowly with time.  These changes ultimately lead to the saturation of the spiral amplitude or quenching of the linear growth (e.g. \cite{sellwood21} and section \ref{sec:disk_evolution}).  
The derivation of the characteristic equation in the plane follows a conventional approach \citep{goldreich79, papaloizou84a}, first combining the linearized Euler equations of motion in cylindrical coordinates with Poisson's equation to derive the in-plane perturbed motions.
These then are substituted into the continuity equation. Whereas the derivation in the appendix of \citetalias{meidt24} considered the 3D continuity equation, here we inspect the 2D version, making our approach most similar to \citet{lin64} and \citet{bertin89a}.

Unlike some earlier usages of the hydrodynamical approximation to study spirals as disk modes, we remove the assumption that the disk is quasi-stable.
This makes it relevant to keep several terms that we find are influential for growth and decay below.
Terms arising with azimuthal forces in particular make it possible to treat resonant and non-resonant amplification with one expression.  These azimuthal terms are neglected in the WKB approximation, but become relevant for short-wave open spirals with $kR\sim m$  \citepalias{meidt24} and increasingly relevant even for small $m$ as $k$ decreases.   

\subsection{The hydrodynamical view of spiral amplification}
Using the linearized equations of motion, in \citetalias{meidt24} we derived a (depressed) cubic characteristic equation for $(\omega-m\Omega)$ with two terms (a linear term and a constant term), each of which dominates the solution to the equation depending on the value of $Q$:  
\begin{eqnarray}
(\omega_e-m\Omega)^3&=&\nonumber\\
(\omega_e-m\Omega)\Big[&\kappa&^2+s_0^2\left(k_e^2+\frac{m^2}{R^2}-\frac{k_e}{R}\left(\frac{\partial \ln\Sigma\Delta}{\partial \ln R}+\frac{1}{R}\right)\right)\Big]\nonumber\\
&+&i s_0^2k_e\Bigg(\frac{2Am}{R}-\frac{\partial(\omega_e-m\Omega)}{\partial R}\Bigg)\nonumber\\
&+&s_0^2\frac{m\Omega}{R}\Bigg[\frac{\partial \ln \Sigma}{\partial \ln R}+\frac{\partial \ln \Omega}{\partial \ln R}-\frac{\partial \ln\Delta}{\partial \ln R}\Bigg]\label{eq:fullchareqn}
\end{eqnarray}
in terms of $\omega_e=\omega+\dot{k}R$ and $k_e=k-i T_1/r$ and
\begin{equation}
s_0^2=-2\pi \Sigma\left(\frac{e^{-i\delta}}{k_{e,long}}\left[1+\frac{\Sigma_b}{\Sigma_{1,m}}\frac{k_{e,long}e^{-i(\delta_{\rm b}-\delta)}}{k_{b,e,long}}\right]-\frac{1}{k_{J,2D}}\right) \label{eq:s0sqlong}
\end{equation}
with $k_{J,2D}=(2\pi G\Sigma/\sigma^2)$.  

Here $\Sigma$ is the unperturbed axisymmetric disk surface density, $\Omega=V_c/R$ is the disk angular velocity, $\kappa$ is the radial epicyclic frequency, Oort $A=-1/2d\Omega/dR$ and $\Delta=\kappa^2-(\omega_e-m\Omega)^2$.
In eq. (\ref{eq:s0sqlong}) 
\begin{equation}
\Sigma_{1,m}=\int_{-\infty}^\infty \rho_{1,m} dz
\end{equation}
and $\Sigma_b$ represents the surface density of an external (or background) potential perturbation with effective wavenumber $k_{b,e,long}$ and potential-density offset $\delta_{b}$, applying the same convention as used for the self-gravity perturbation in section \ref{sec:wkb}.  

The factor $s_0^2$ in eq. (\ref{eq:s0sqlong}) for any mode $m$ can include an additional set of nonlinear terms
\begin{equation}
s_{0,nl}^2=  -\frac{\Sigma_{1,m'}}{\Sigma_{1,m}}\left(\frac{2\pi\Sigma_{1,m''}e^{-i\delta_{m''}}}{k_{e,long,m''}}-\sigma^2 \right)\label{eq:s0sqlongNL}
\end{equation}
that track the contribution from any modes $m'$ and $m''$ that satisfy $m=m'+m''$ or $m=m'-m''$ (see e.g. \citealt{sygnet88} and \citealt{tagger87,masset97}).  This contribution is relevant when the perturbed surface densities $\Sigma_{1,m}$, $\Sigma_{1,m'}$ and $\Sigma_{1,m''}$ become comparable to $\Sigma_0$ and $k_{e,long,m''}\lesssim k_{e,long,m}$.  

For $Q\lesssim 1$, the characteristic equation assumes the form of the conventional quadratic tight-winding dispersion relation (in the  WKB approximation), yielding the well-known conditions for growing spirals, 
\begin{equation}
(\omega_e-m\Omega)^2=
s_0^2\left(k_e^2+\frac{m^2}{R^2}-\frac{k_e}{R}\left(\frac{\partial \ln \Sigma_0}{\partial \ln R}+\frac{\partial \ln\Delta}{\partial \ln R}+\frac{1}{R}\right)\right).
\label{eq:conventional}
\end{equation}
This expression is outside the tight winding limit, even though it is typical to still assume $k\gg m/R$.
This is close to the relation derived by \cite{bertin89a}.

In the regime $1\lesssim Q\lesssim 6$, solutions to the characteristic equation are dominated by the constant term in eq. (\ref{eq:fullchareqn}) and obey 
\begin{eqnarray}
(\omega_e-m\Omega)^3&\approx&\label{eq:long}\\
&+&i s_0^2k_e\Bigg(\frac{2Am}{R}-\frac{\partial(\omega_e-m\Omega)}{\partial R}\Bigg)\nonumber\\
&+&s_0^2\frac{m\Omega}{R}\Bigg[\frac{\partial \ln \Sigma_0}{\partial \ln R}+\frac{\partial \ln \Omega}{\partial \ln R}-\frac{\partial \ln\Delta}{\partial \ln R}\Bigg]\nonumber
\end{eqnarray}

These terms are prominent outside the tight winding limit with $k\sim m/R$ and represent the impact of the donkey effect that becomes activated in shearing disks in the presence of azimuthal (non-axisymmetric) forcing (\citetalias{lynden-bell72}; \citetalias{meidt24}).  
The first term on the right becomes $i s_0^2k_e 2Am/R$ at corotation and is negligible elsewhere \citepalias{van-der-wel25}.
In the second term on the right, the second two factors originate with the radial gradient in the perturbed radial velocity while the first term in square brackets represents the radial advection of the wave.

\subsubsection{A brief summary of paths to spiral growth}
As discussed in \citetalias{meidt24} and \citetalias{van-der-wel25}, eq. (\ref{eq:long}) offers a powerful, compact expression that captures several related forms of wave amplification.
(For growth, the imaginary component of solutions must be positive, i.e. the growth rate $\beta>0$.) The first term on the right, which we call the 'corotation amplification term', depicts both swing-amplification and its cousin corotation amplification.
(When this term dominates in the short-wave regime it signifies a purely imaginary solution that pulls growth to corotation.) Swing-amplification and corotation amplification both fundamentally leverage the same physical mechanism, namely inverse Landau damping \citepalias{lynden-bell72}.
That inverse damping taps into the 'donkey effect' in which orbiting stars or parcels of gas respond to a perturbation at corotation as if responding to a negative mass, moving forward when pulled back (and vice versa).
We find this description from the stellar dynamical realm to be a very intuitive way of describing spiral amplification even in a fluid description, although other more wave-like pictures can also be invoked (see below).

The second term (or the long-wave term) plays multiple functions.
Perhaps its best known role is in regulating the properties of waves launched from corotation, as in the WASER mechanism \citep{mark74, mark76} and  the corotation amplification picture of \citep{papaloizou89}, as discussed more in \S \ref{sec:exponential_caverna}.
This role emerges in eq. (\ref{eq:long}) clearly by rewriting the long-wave term in terms of the gradient in potential vorticity $\Sigma\Omega/(2\kappa^2)$.
When such a gradient is present at corotation, it leads to over-reflection, allowing the outgoing wave to amplify more than it would in the absence of the vorticity gradient.

Aside from this role, the second, long-wave term depicts two other avenues to amplification, independently of the 'corotation amplification term'.
In fluid dynamical calculations, the first case is typically associated with sharp gradients in potential vorticity.
These lead to the amplification of Rossby waves at corotation, as described by \citet{lovelace78} specifically in the context of galactic disks (and \citealt{lovelace99} and \citealt{li00} in planetary and accretion disks, respectively).
The groove modes discussed by \citet{sellwood91, sellwood02, sellwood12} in stellar half-mass Mestel disks amplify similarly;
just as in the case of corotation amplification, the amplified waves are the product of a 'negative mass' instability.

Milder gradients in potential vorticity can lead to the second type of 'long-wave' amplification that is exclusively non-resonant.
As we describe later in \S \ref{sec:gdi_def}, this gradient, which we prefer to call a gradient in the inventory of donkeys (or GDI), is related to what \citet{polyachenko19} refers to as the Lynden-Bell derivative of the distribution function.
In some scenarios, the sign of the GDI yields a stable, steady wave inside or outside corotation (see next section).
When the non-resonant structure is amplifying, the growth is necessarily short-lived as a result of the non-resonant heating produced.
This quality seems to have led to an undervaluing of this amplification relative to the fast growing groove modes or Rossby wave instabilities that are capable of greater maturity\footnote{This amplification is arguably linked to the non-resonant orbital diffusion found to be influential by \citet{fouvry15, heyvaerts17}}.
However, as we describe below, typical galactic disks frame a mild GDI for the duration of their disk-like structure, making non-resonant spirals a rich source of secular evolution.

Below we will summarize the basic qualities of the solutions to eq. (\ref{eq:long}) exclusively with mode-like spiral formation in mind (whereby $\dot{k}=0$) at a stage in which the self-gravity response dominates the initial seeding perturbation and thus assume $\Sigma_1\gg \Sigma_{b,e}$.
That is, we use eq. (\ref{eq:long}) to depict the self-gravity 'dressing'.  We will also explicitly ignore the contribution from nonlinear coupling given by eq. (\ref{eq:s0sqlongNL}), although its role is discussed again later in $\S$ \ref{sec:nonlinearev}.  

\begin{figure*}[t]
\begin{center}
\begin{tabular}{c}
\includegraphics[width=.95\linewidth]{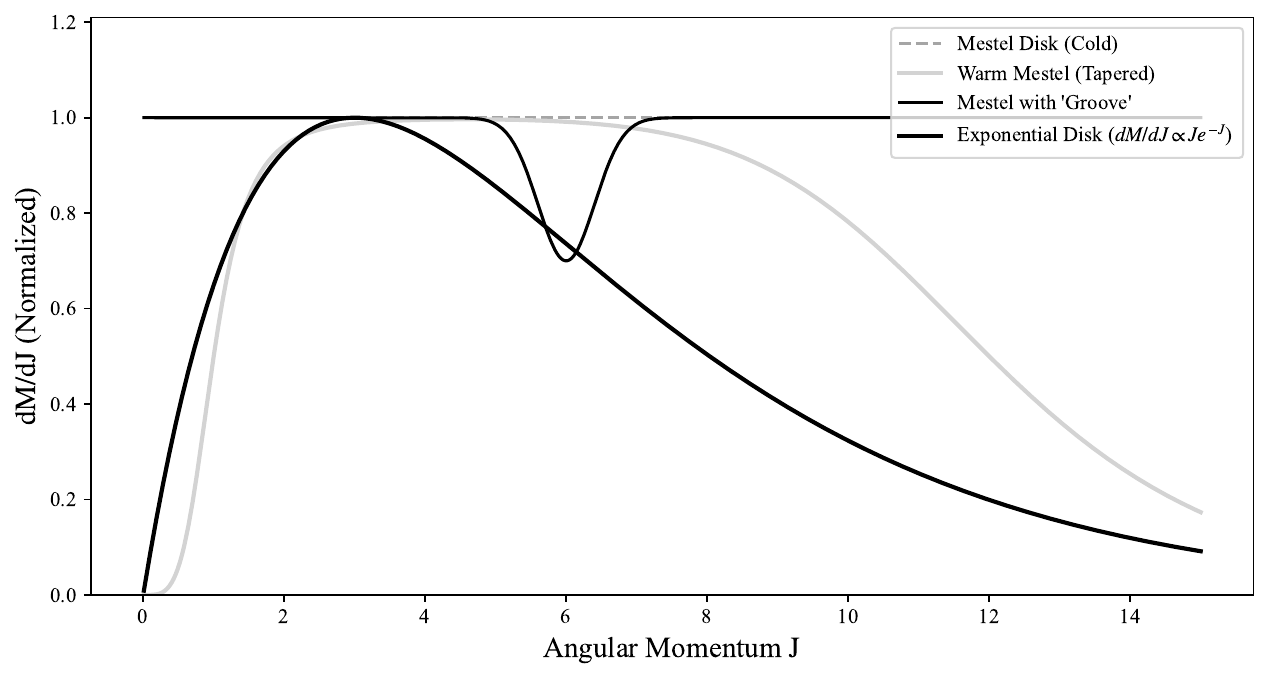}
\end{tabular}
\end{center}
\caption{The normalized mass per unit angular momentum, $dM/dJ$, for four representative galactic disk models. The dashed gray line shows an example of a cold Mestel disk, for which $dM/dJ$ is constant (equal to $V_c/G$, in terms of the circular velocity  $V_c$). The solid gray line shows a warm Mestel disk with tapering both at small and large $J$ to represent a finite disk with a truncated inner core and outer boundary. The thin black line highlights an example of a grooved Mestel disk, with a localized deficiency, or "groove," centered at $J = 6.0$ (see e.g. \cite{sellwood91,sellwood22}). The thick black line illustrates an exponential disk embedded in a potential with a flat rotation curve, which exhibits a characteristic $J \exp(-J)$ profile. 
}
\label{fig:illustration}
\end{figure*}

\subsection{Amplification in different systems}\label{sec:amp_systems}
The form of eq. (\ref{eq:long}) yields complex solutions that correspond to resonant and non-resonant growth and decay, depending on the properties of the wave and the conditions in the disk, as discussed in detail in \citetalias{meidt24} and \citetalias{van-der-wel25}.
In that earlier work, resonant and non-resonant (large-scale) amplification were described independently by separately examining solutions in the short- and long-wave regimes, respectively.
Below the description instead emphasizes the influence of conditions in the disk on where the resonant and non-resonant growth occurs.
This makes it easier to relate the view offered by a hydrodynamical approach to the picture that has been assembled so far with numerical simulations and stellar dynamical calculations.

With this goal in mind, below we discuss spiral formation in several well-studied scenarios.
For this discussion, our goal is to examine the impact of disk properties on growth or steadiness, rather than the influence of the spiral's own properties (which we address in \S \ref{sec:instantaneous}).
For now, then, we implicitly assume that the spiral perturbation has properties most conducive to growth, i.e. that self-gravitational forces exceed pressure forces and $s_0^2<0$.
We thus do not consider acoustic waves (for which pressure dominates).
For this optimal growth scenario we also assume that $T_1\lesssim 0$, making the imaginary component of $s_0^2$ less influential than the real part.
We postpone a discussion of phase variations originating with $s_0^2$ to later.

\subsubsection{The stability of cold Mestel disks}
Mestel disks are a favored test-bed for building intuition about spiral arm formation, evolution and recurrence.
Cold Mestel disks are famously stable against large-scale spiral formation.  This is depicted by eq. (\ref{eq:long}) through the cancelling of the long-wave terms; in Mestel disks,  $d\ln \Sigma/d\ln R=d\ln \Omega/d\ln R=-1$.
Although disks like these yield no long-wave growth, according to eq. (\ref{eq:long}), we can still expect short-wave $kR\gg1$ features to grow if $m$ is also non-zero (such that the linear term in eq. (\ref{eq:long}) is non-zero).
The growth rate in this case is fastest for 'open' spirals for which $m$ is comparable to $kR$ \citepalias{meidt24}.
When $kR\gg m$ in the tight-winding limit, on the other hand, growth is suppressed unless $Q<1$ in which case the characteristic equation becomes the normal tight-winding Lin-Shu dispersion relation \citepalias[see][]{meidt24, van-der-wel25}.

\subsubsection{Large-scale Spirals in warm Mestel disks}
Alongside the well-known global stability of Mestel disks \citep{zang76,toomre77,toomre81}, Mestel disks that are dynamically warm and have a central `cut-out' or $Q$-barrier' exhibit the suprising capability of forming large-scale $m=2$ patterns, as shown by \citet{zang76}.
Two factors in eq. (\ref{eq:long}) combine to depict the sort of global, large-scale spiral in this case.
First, the warmth cuts off the growth of large $k$ and large $m$ perturbations so that corotation amplification favors low $k$ and $m$ perturbations.
Second, the warmth makes the `Dehnen drift' relevant and shifts the long-wave term away from zero.
We use the unperturbed equations of motion for our underlying axisymmetric disk to rewrite $\Omega=V_{\theta}/R$ explicitly as a function of the velocity dispersion $\sigma$, i.e. 
\begin{equation}
\Omega^2\approx\Omega_c^2-\frac{\sigma_r^2}{R^2}
\end{equation} 
with the centripetal force set by the gravitational force and the pressure force (in the hydrodynamical approximation).
In this case the long wave term becomes non-zero, i.e. $d\ln \Sigma/d\ln R-d\ln \Omega/d\ln R\approx\ln \Omega/d\ln R (1+\sigma^2/V_c^2)$, yielding a steady (non-growing) loss-less spiral inside corotation fed by corotation amplification as long as $d\ln \Omega/d\ln R<0$, which is the case in Mestel disks and most galactic disks.

\subsubsection{Rossby waves and Groove modes in full or half-mass Mestel disks}
Using tailored numerical simulations, Sellwood translated the link between the warmth in a Mestel disk and spiral-forming ability into a general condition on the distribution function of stars \citep[e.g.,][]{sellwood89, sellwood12, sellwood14}: Mestel distributions that are 'grooved' -- in the sense that warmth leads to a deficit of stars centered on a given angular momentum -- characteristically become unstable to spirals.
Such a groove could be incorporated into eq. (\ref{eq:long}) either as 'warmth' (as described above) or as an local gradient in the density $d\Sigma_{\rm groove}/dR$ proportional to the width of the desired groove.
Grooves in stellar disks are analogous to the sharp features in potential vorticity that \citet{lovelace99} determined produce the Rossby wave instability in fluid disks.
Both processes leverage the 'negative mass' response characteristic of donkey behavior in differentially rotating galactic disks, and both types of features saturate through the redistributon of mass across the groove (discussed more in Section \ref{sec:disk_evolution}).
The groove mode phenomenon is a key fixture in modern theory as it gives rise to recurrence and sustained spiral structure.
The spiral groove modes create new grooves at the spiral's ILR and/or OLR where the wave heats the disk, establishing the method of spiral recurrence \citep{sellwood89,sellwood91,sellwood12,sellwood14,sridhar19}. 

\subsection{The amplification of Spirals in Exponential disks}\label{sec:exponential_caverna}
\subsubsection{Cavernae}
Prototypical spiral galaxies often lack the same degree of self-consistency as warm or cold Mestel disks in the sense that the stellar disk component does not generate the potential and circular motion (or is not proportional to the dominant mass component, as in the case of the half-mass Mestel disk).
Instead we typically have a situation where an approximately exponential distribution of stars is embedded in a dominant dark matter (DM) halo with an approximately constant velocity curve.
In this state, the stellar disk can be deficient in stars over a wide range in angular momentum, making large parts of the disk act like a wide (rather than sharp) groove.

Consider, first, the equal distribution of stars characteristic of a perfectly cold Mestel disk with $dM/dj=$ constant, as plotted in Figure \ref{fig:illustration}.  
In comparison, exponential disks embedded in a background (DM) potential, like warm Mestel disks, contain a deficit of low angular momentum stars inside $R\sim R_e$.  
They also `miss' stars at high angular momentum. (In an exponential disk, $dM/dj=\int f(E,J),dE\propto \textrm{exp}\big[J/(RV_c)\big]$.)  In a hollow in the disk distribution like this—which we will refer to as a `caverna'—the long-wave terms are non-zero, tracking instead a natural, mild gradient in the potential vorticity.
In this case, eq. (\ref{eq:long}) predicts the formation of spirals that work to transport angular momentum outward and remove the deficit.

The density gradient that feeds growth through the donkey effect (inverse Landau damping) in an exponential caverna is much milder than the strong gradients leveraged in the rapid, strong growth of groove modes and Rossby waves, as well as edge modes.
Such a mild gradient is essentially imperceptible to high-$k$ spirals, but it can be substantial when the spiral is long and open (low $k$), sufficient for donkey behavior to produce a net transport of angular momentum outward.

\subsubsection{Relation to the donkey effect}
In modern spiral theory, donkey behavior leads to a `dressing' response that is typically strongest at corotation (or near sharp density gradients) where it is sustained (\citetalias{lynden-bell72}; see also \citealt{julian66} and \citealt{fouvry15} for the dressed response), but differential rotation can turn stars into donkeys anywhere.
Away from corotation, the time stars spend responding to a spiral is characteristically short and orbiting stars (or gas parcels) only temporarily execute supporting donkey behavior.
But because they do so one after the other, they share the burden of supporting the spiral.
Add to this a modest negative density gradient and the result is a net gain of angular momentum inside corotation that settles donkeys downstream in a position to support the (trailing) spiral.
The result is spiral growth in many cases (see also the calculations and discussion in section \ref{sec:heating}).
In other words, with the right radial distribution, donkeys are present in enough numbers even away from corotation to help each other assemble a large-scale pattern and transport angular momentum outward.

Because these participating stars are temporarily joined in their donkey-like behavior to bridge the caverna, we refer to them as `asellae nexae cavernae' (or `nexae cavernae' for short). The large-scale, emergent pattern that they collectively build is what we define as a `spiral nexum' (plural `spiral nexa'). We apply this naming convention primarily to distinguish these widespread features in exponential-like disks from the more localized groove modes that have been studied primarily in Mestel disks, though a spiral nexum can ultimately develop from the union of diverse components.

In this light, we view the gradient in potential vorticity as a ``gradient in the inventory of donkeys" (GDI) or (more exactly) as a phase-space distribution gradient ($\partial f / \partial J$), as we discuss in more detail later in \S \ref{sec:gdi_def}.

In the following sections (\S\ref{sec:comp1} - \S\ref{sec:comp3}) we describe in more detail the conditions in exponential caverna that lead to different kinds of spirals and spiral growth.
This includes resonant dressing,  long-wave non-resonant structures that can be both steady and non-steady, and groove modes stimulated by the secular changes brought about by steady structures.
The contributions of each component are sensitive to the disk properties and they will evolve with different timescales, as we ultimately show in \S \ref{sec:disk_evolution}.

\subsection{Spiral nexa Components I:  resonant dressing}
\label{sec:comp1}
Through the corotation amplification term, perturbations are capable of supporting resonant dressing essentially anywhere.
The amplification of the longest of these (small $k$)  is boosted with the addition of the negative GDI (potential vorticity) gradient.
(For higher $k$, the change to the amplification rate in the presence of non-zero GDI is negligible.) As described in section \ref{sec:instantaneous} the perturbations that grow the fastest are those that maximize the gravitational forcing without exceeding $k_J$.
As mentioned above, the GDI will tend to give spirals a boost in their corotation amplification (\citealt{papaloizou84, mark76}; \citetalias{van-der-wel25}).

\subsection{Spiral nexa Components II: steady and non-steady non-resonant structures}
\label{sec:comp2}
Away from resonances, the GDI in typical galactic disks can either help spirals grow or it can stabilize them.
To see non-resonant structures away from corotation, we look to solutions to eq. (\ref{eq:long}) when the long-wave term dominates.
(Away from corotation, the corotation amplification term goes to zero).
Solutions can correspond to either growth and decay or stability either inside or outside corotation depending on the sign of the long-wave term term.
For example,  if $s_0^2<0$ (self-gravity dominant over pressure), then growth inside corotation requires the long-wave term to be negative, or 
\begin{equation}\label{eq:longin}
\frac{d \ln \Sigma\Omega}{d \ln R}-\frac{d \ln \kappa^2}{d\ln R}-\frac{d \ln\left(1-\frac{(\omega-m\Omega)^2}{\kappa^2}\right)}{d \ln R}<0.
\end{equation}
Note that the third term on the left stays smaller than the second term essentially everywhere except near the Lindblad resonances, where $(\omega-m\Omega)=\pm\kappa$, at which point solutions to eq (\ref{eq:longin}) are undefined.  This defines the maximum outer extent of the spiral.  

Between the CR and the ILR or OLR, the left hand side of eq. (\ref{eq:longin}) can be well approximated by the gradient in the potential vorticity or GDI and so we have growth or stability depending on whether $\Sigma\Omega/\kappa^2$ is increasing or decreasing with radius.
Specifically, when the GDI is negative (long-wave term is positive), these solutions correspond to growth and decay inside corotation and steadiness outside corotation.
Instead, when the GDI is positive (long-wave term is negative), we find growth and decay outside corotation and steadiness inside corotation.
Galactic disks can have both positive and negative GDI regimes, although some can also have a negative GDI everywhere, suggesting a diversity of spirals are possible.
\begin{figure}[t]
\begin{center}
\begin{tabular}{c}
\includegraphics[width=.995\linewidth]{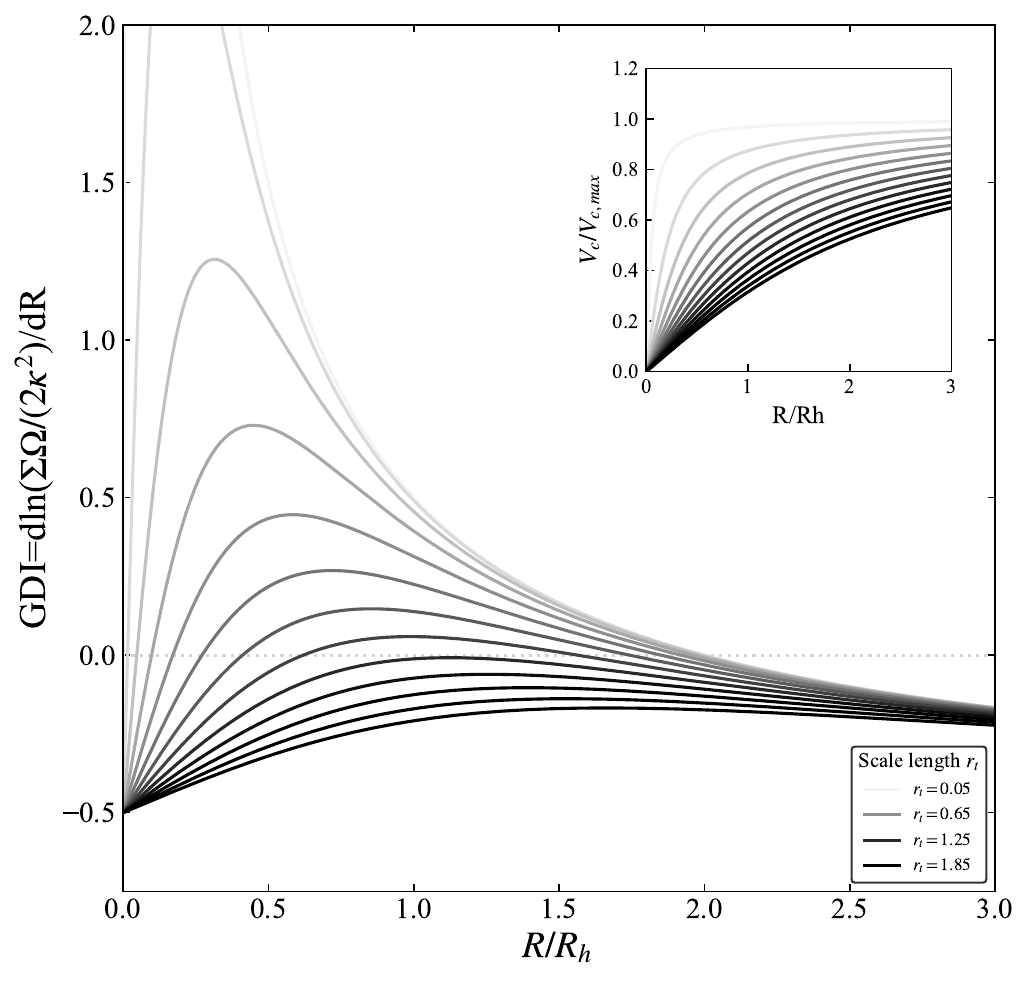}
\end{tabular}
\end{center}
\caption{Illustration of the variation in GDI (gradient in the inventory of donkeys) $d\ln\Sigma\Omega/(2\kappa^2)/dR$ as a function of $R/R_h$ for exponential disks with scale length $R_h$ embedded in a potential with circular velocity curve $V_c=V_{c,max}(2/\pi)\arctan(R/r_t)$, plotted for reference in the inset in the top right. The different curves correspond to different values of $r_t$, the transition radius where the rotation curve switches from rising to flat.  The plotted values span the range $0.05~\textrm{kpc}\leq r_t \leq 1.85~\textrm{kpc}$ (from low to high, gray to black). 
A GDI of zero is marked with a horizontal line.  When the curves sit below (above) zero, the GDI $d\ln\Sigma\Omega/(2\kappa^2)/dR$ is negative (positive), implying that spiral waves are non-steady (steady). Note how extended, slowly rising rotation curves (large $r_t$) maintain a negative GDI everywhere, while steeply rising curves (small $r_t$) open a zone of positive GDI in the inner disk. 
}
\label{fig:GDIplot}
\end{figure}

\subsubsection{The architecture of caverna -- donkeys at work}\label{sec:caverna_architecture}
The behavior of the long-wave term defines the architecture of the caverna and the nature, properties and number of spirals at any given time.
This is depicted in Figure \ref{fig:GDIplot}. Galaxies with slowly rising rotation curves have a GDI that is mostly negative everyhwere.
But when the rotation curve rises more steeply (so that the rotation curve overall varies like $\sim R$), a zone of positive vorticity opens up inside $R=R_e$, opening with it the possibility of steadiness inside corotation.
Note that this behavior is opposite to what is predicted from a PV gradient in the absence of self-gravity, in the case of acoustic waves.

The steadiness or non-steadiness can be understood in terms of how the donkey effect supports waves.
For waves lagging the disk ($\Omega_p<\Omega$), a negative GDI signifies there are sufficient donkeys to yield a net positive angular momentum whereby donkeys settle downstream of the trailing spiral and support growth.
Outside corotation where waves overtake the disk ($\Omega_p>\Omega$), on the other hand, the steady solution signifies that there is a precise wave speed (for a given set of spiral properties) at which the wave's ability to collect stars via the donkey effect is outpaced by the wave's advection through the stars, keeping it steady rather than allowing it to grow.
The reverse occurs when the GDI is positive, such as at $R<R_e$.
In this case, the density gradient is still negative allowing the donkey effect to shift angular momentum outwards, but now wave growth is outpaced by the wave's advection of stars.
A greater number of stars moving outward (through a larger density imbalance) are needed than are present and the wave is steady rather than growing in place.

The amplification experienced inside or outside corotation is necessarily transient, given that the non-steady wave will also lead to non-resonant heating.
This is the focus of section \ref{sec:heating}.
Still, at any given moment in time, the spiral nexum in a given disk can consist of a variety of spiral components, some of which are steady and others that are non-steady.
The transition will primarily be marked by the location in the disk where the long-wave term goes to zero (if at all).
For the systems that have extended flat rotation curves, for example, this coincides with $R\approx R_{e}$ or $R\approx R_{e} (1-\sigma^2/V_c^2)^{-1}$ including the Dehnen drift.
The warmer the disk, the larger the steady zone.  The steady spiral (inside $R_e$) and non-steady spiral (outside $R_e$) in this case will necessary have different speeds;
in order that the growing spiral outside $R_e$ is inside corotation, it must rotate slower than the steady spiral inside $R_e$.
In typical disks, the non-steadiness at the lower speed outside $R_e$ will transition back to steadiness at corotation.

In the regime $R>R_e$ (or anywhere the GDI is negative) growth outside corotation requires a positive density gradient.
This is the situation presented by the outer edge of a groove, but it would also occur in the event of an upward bending profile \citep[e.g.,][]{fiteni24}.

\subsubsection{A spiral for every endo and exogenic density profile feature}
The division of a disk into stable or unstable regions is most sensitive to the rotation curve shape.
Within these regions, the architecture of caverna can incorporate secondary features originating with inflections in the disk density distribution.
For each bend in the density distribution (tracing more than one scale length $R_h$), the disks can harbor another spiral.
This includes outer bends that give rise to edge modes \citep[e.g.,][]{sellwood89,fiteni24}.
These inflections and bends can be both endo- and exogenic, formed through interactions and minor mergers or produced by the spirals naturally forming in the disk.
In much the same way \citet{sellwood12} and \citet{sellwood19} have uncovered in the study of recurrent groove modes in Mestel disks, spiral nexa can reshape the cavernae, carving other caverna or relocating inflections in $dM/dj$, thus influencing the spectrum of possible patterns for the next generation of spirals: the steady components have their influence at resonance, while non-steady components can have a more widespread influence, carving features that are milder than grooves.

\subsubsection{The link between Q barriers and steady, `stable' modes}
The non-steady structures depicted in eq. (\ref{eq:longin}) are necessarily short lived;
their non-resonant growth leads to exchanges of angular momentum in place and consequently to disk heating (see \citetalias{van-der-wel25} and section \ref{sec:heating}).
The more classical, slowly evolving structures are the steady waves depicted by eq (\ref{eq:long}).
These map onto the global 'modes' identified by \citet{bertin97}  that efficiently transport angular momentum between the ILR and CR \citep{goldreich79, binney08}.
Our approach re-emphasizes that these steady spirals are intimately linked to both dynamical warmth and the shape of the rotation curve, and in particular with the presence of a steeply rising rotation curve.

Both of these qualities are present under the same circumstances as a `Q barrier'.
In the conventional view, the `Q barrier' is envisioned as a structure that supports the reflection of an ingoing wave back outwards to corotation that allows a standing wave to develop.
We prefer to think of it as opening a regime in the disk in which the inventory of donkeys is not able to lead to the coordinated growth of a pattern.
The rapid variation in $\Omega$ at small radii requires a larger density gradient than present for donkeys to efficiently settle downstream in the trailing spiral.
Instead, the rate at which 'donkeys' collect in lagging waves (inside corotation) matches the pace of the wave's advection and the wave can only propagate, rather than grow.

This behavior has interesting implications for how the spiral (and bar) structures hosted by a given galaxy evolve over time.
Galaxies may start off with slowly rising rotation curves and as mass moves inward, the rotation curve steepens and quickly flattens, which has the effect of stabilizing spirals and/or bars, making them steady. 

\subsubsection{The GDI as the gradient in the distribution function: Global, mild GDI vs. localized, sharp GDI }\label{sec:gdi_def}
The influential role of the GDI (or potential vorticity gradient) that appears in our fluid dynamical calculations is arguably best appreciated when linked directly to $\partial f/\partial J$, which \citetalias{lynden-bell72} identified as the source of secular changes at resonance, or what \citet{polyachenko19} call the Lynden-Bell derivative of the distribution function.
Writing
\begin{equation}
2\pi \Sigma RdR=(2\pi)^2dL \int_0^\infty f(L,J_R) dJ_R  
\end{equation}
for a small mass element, then  
\begin{equation}
\int_0^\infty f(L,J_R) dJ_R=\frac{1}{2\pi}\frac{2\Omega\Sigma}{\kappa^2}
\end{equation}
or 
\begin{equation}
\int_0^\infty\frac{\partial f}{\partial L}dJ_R\propto \frac{\partial }{\partial R}\frac{2\Omega\Sigma}{\kappa^2}
\end{equation}  
using that $dL/dR=2\kappa^2/(2\Omega)$.
This proportionality underscores the significance of the mild GDI identified here in relation to the role conventionally played by GDI in modern theory, namely as a 'source term' for the exchange of angular momentum between the disk and the spiral wave at resonance or a sharp feature. In the framework of \citetalias{lynden-bell72}, for example, the gradient in the density of stars in action space $\partial f/\partial L$ determines the net torque and allows the wave to "trap" or scatter particles, extracting energy from the disk to fuel wave growth.

This same functionality is behind the concept of groove modes developed by \citet{sellwood89} and  \citet{sellwood12}.
A sharp, narrow "groove" or deficit in the distribution function (a localized spike in $\partial f/\partial L$) triggers a powerful instability as the disk attempts to "fill" the phase-space gap through resonant transport.
In parallel, the radial gradient of the inverse potential vorticity is key for angular momentum transfer and wave launching at corotation in fluid disks \citep{mark76, goldreich79}.
Localized extrema or sharp gradients in the potential vorticity are seen as a necessary condition for the RWI \citep{papaloizou84, lovelace99}.

The difference between the way those earlier works envision the role of the GDI, or distribution function gradient, and the way it functions here is related to the spatial extent of the zone where the GDI has its influence. 
The focus on the GDI at resonances (or in the form of extrema) in prior literature arguably stems from an underlying interest in describing the growth and evolution of steady spirals in stable disks.
Steady spirals exchange energy and angular momentum with the disk primarily at resonances \citepalias{lynden-bell72}, making these the most relevant locations.
But the behavior of the GDI (and $\partial f/\partial J$) becomes relevant away from resonance, especially when we relax the assumption of quasi-steadiness to describe non-steady spirals.
As they propagate, these transient spirals do not conserve wave action  and preferentially exchange with the disk away from resonance.

In this light, the donkey effect and $df/dJ$ takes on a more pervasive, global role than envisioned in the steady calculations of \citetalias{lynden-bell72}; it is also fundamental for describing the emergence of non-resonantly growing, non-steady spirals.  
This aspect has been not been widely valued, however, given that the short lifetimes of individual non-steady spirals has traditionally made their steady counterparts the primary focus for studying secular evolution.  The clear view of the conditions responsible for non-resonant excitation in this work, especially their long duration, on the other hand, underlines the vital role these collective features can have for the evolution of global disk properties.  This compliments the perspective built from stellar dynamical calculations, which show that non-resonant secular evolution constitutes a vital, continuous channel for structural relaxation, 
orbital diffusion, and star-disc scattering in a diversity of self-gravitating stellar systems 
\citep[e.g.,][]{fouvry15, heyvaerts17, roule25}.
\begin{figure}[t]
\begin{center}
\begin{tabular}{c}
\includegraphics[width=.995\linewidth]{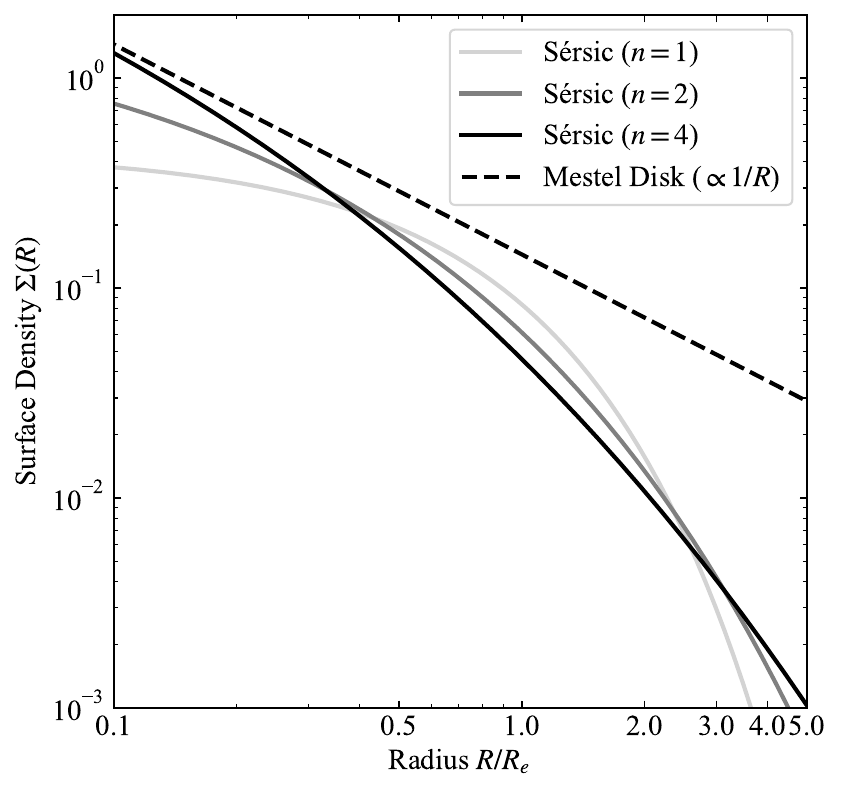}
\end{tabular}
\end{center}
\caption{Illustration of the four representative galactic radial surface density profiles $\Sigma(R)$. The light gray line shows an $n=1$ S\'ersic profile corresponding to an exponential disk (typical of late-type spiral galaxies).  The gray and black lines show an $n=2$ profile and an $n=4$ de Vaucouleurs profile, respectively (typical of early-type fast- and slow-rotators).  The black dashed line shows a Mestel disk with $\Sigma(R) \propto 1/R$.  All profiles are normalized to have the same integrated mass within $10 R_e$ (in terms of the effective radius $R_e$).  The three S\'ersic profiles in this illustration have the same effective radius $R_e$. As spiral nexa redistribute angular momentum and heat the disk, they drive the system's structural evolution from a cold exponential ($n=1$) toward the theoretical maximum-entropy Mestel state, frequently stalling at the more centrally compacted, dynamically warm morphologies ($n \ge 2$). 
}
\label{fig:profiles}
\end{figure}
\subsection{Spiral nexa Components III: groove modes}
\label{sec:comp3}
The secular changes brought about by steady spirals in exponential disks contribute a third type of feature in exponential disks, namely groove modes \citep{sellwood89, sellwood14, sellwood19}.
When present, groove modes make a strong contribution to the morphological appearance of the disk and have a dominant role in radial migration.
In the arrangement described in \S \ref{sec:caverna_architecture}, we expect that these would mostly appear in the steady-spiral regime inside $R_e$, but they could also potentially be triggered in the zone hosting non-resonantly growing spirals $R>R_e$, depending on the arrangement of resonances and where grooves are carved.
We refer the reader to \citep{sellwood14a} and \citep{sellwood22a} for reviews of how these features form and evolve.

Here we will try to sketch (incompletely) how groove modes appear in comparison with the spirals built as nexae in caverna.
As discussed in \S \ref{sec:exponential_caverna}, spiral nexa can be thought of as milder, one-sided cousins of groove modes that are associated with the natural deficit in the disk distribution function.
But there are subtle differences between the two kinds of spirals.
For one, caverna form through broader, rather than strictly resonant, secular changes, whereas grooves tend to be narrow features, with each spiral of a given speed mostly confined to within its groove.
The exponential caverna is more widespread and one single spiral can extend over a wide area.
(Multiple grooves, which are a prediction that accompanies grooves in the groove picture, could produce a spiral that spans a larger radial range in the disk, so radial extent may not be sufficient to distinguish groove modes from spiral nexa.) 

Differences in spatial extent lead to differences in the next generation of spirals that each tends to produce.
Whereas groove modes and steady spirals produce new grooves at resonances \citep{sellwood19}, the widespread changes in angular momentum and heating produced by non-steady spiral nexa produce caves and inflections in the disk that are wide rather than narrow.
These changes allow spiral nexa to benefit from the same basic mechanism as groove modes for their recurrence: one spiral nexum produces changes to the disk that carve a new caverna, these are just less narrow than produced by grooves.

An important difference, however, is that spiral nexa produce non-resonant heating that is as influential as changes in angular momentum at corotation.
So while the reshaping of the disk distribution across resonances or grooves is the primary limit to groove modes, heating contributes to the lifetimes of individual spiral nexa episodes (see $\S$ \ref{sec:disk_evolution}).

The global spiral morphology in a given galactic disk at any given time can be built from both spiral nexa and groove modes.
Proximity to resonance (i.e. established with pattern speed measurement) would be one clear way to identify which type of structure is present at any given location.
Differences in the phase space morphology produced by steady and non-steady spirals \citep[i.e.,][]{tremaine23} provides another a powerful way to observationally distinguish between the two, at least in the MW, although this is not something that can be addressed with the fluid dynamical approach in this work.

\begin{figure*}[th]
\begin{center}
\begin{tabular}{c}
\includegraphics[width=.95\linewidth]{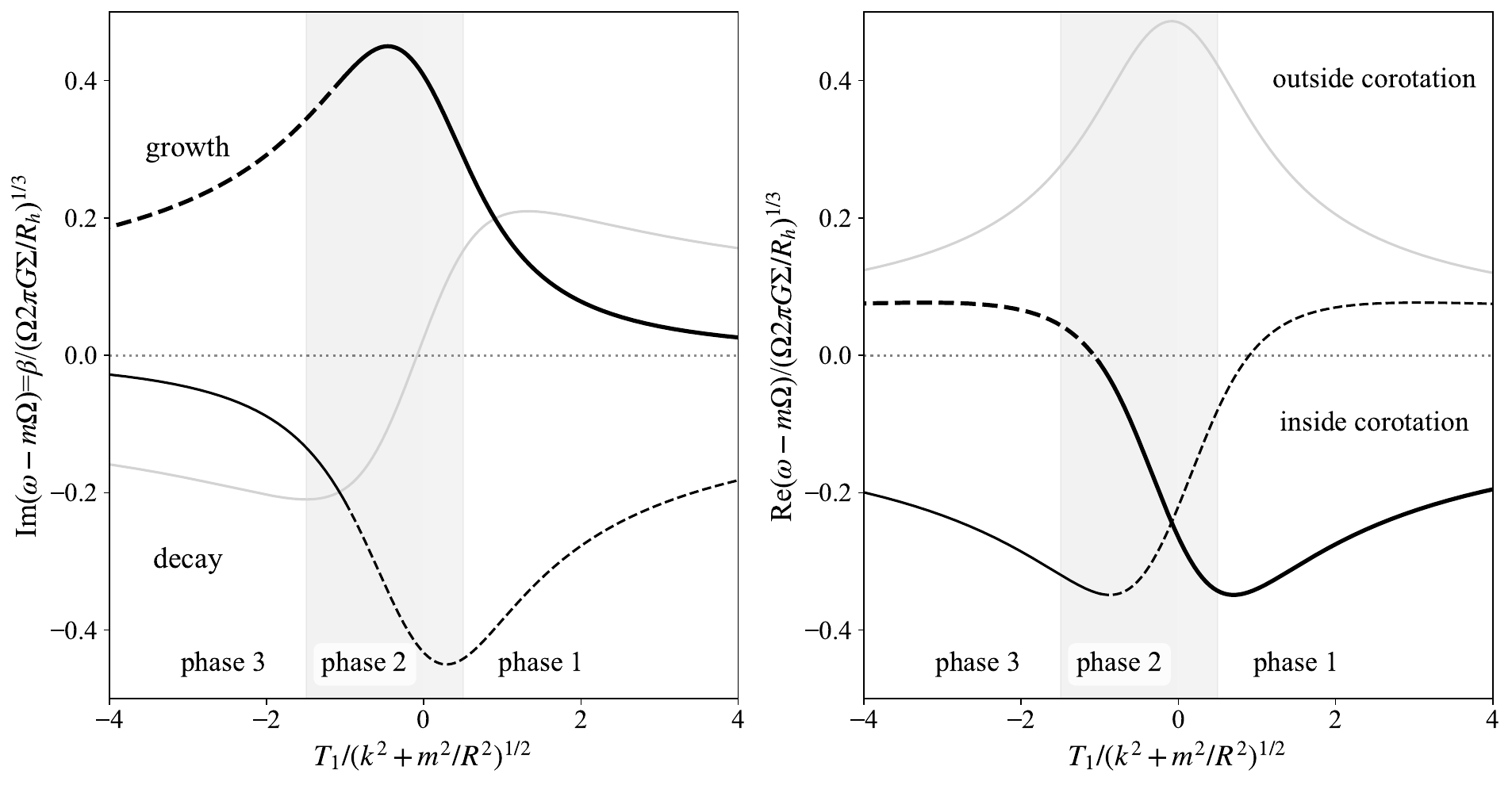}
\end{tabular}
\end{center}
\caption{Solutions of the cubic characteristic relation in eq. (\ref{eq:long}) for $\omega - m\Omega$ away from corotation, plotted as a function of $T_1$ normalized by $k_t=\sqrt{k^2+m^2/R^2}$. The imaginary (left) and real (right) parts of the three solutions are shown.  The thin gray line corresponds to a solution that is fully outside corotation for the entire plotted range in $T_1$.  The thick and thin black solid lines correspond to growing and decaying solutions that are inside corotation for at least part of the plotted range.  The linestyle for the thick black line changes from solid to dashed where the solution changes from inside to outside corotation, when $T_1\approx\sqrt{k^2+m^2/R^2}$.  The dashing for the decaying solution inside corotation switches from dashed to solid at the same location.  The light gray shading in both panels highlights phase 2 when the growth inside corotation is maximized.  Note how the primary growing solution (thick line) traces the full lifecycle of a spiral nexum: initiating in Phase 1, accelerating to peak growth in Phase 2 as the donkey effect strengthens, and ultimately transitioning to decay in Phase 3 as the steepening amplitude gradient ($|T_1|$) weakens self-gravity.}
\label{fig:solutions}
\end{figure*}

\subsection{The ultimate fate of cavernae}
Whether in the form of spiral nexa or groove modes (or a combination of both), spirals ultimately act to remove the exponential caverna altogether \citepalias{lynden-bell72}, leaving a maximum entropy, minimum angular momentum core tightly packed inside an extended, low density high angular momentum envelope.
Until that maximum entropy state is reached, eq. (\ref{eq:long}) suggests that the temporary 'caverna' state may be a preferred arrangement in which spirals are continuously present and active (see discussion in Section \ref{sec:disk_evolution}).

According to eq. (\ref{eq:long}), the optimal `end state', in which spiral nexa are no longer needed to redistribute angular momentum, is one with a Mestel-like equal distribution of stars at all angular momentum.
The singularity in the Mestel disk gives it a higher entropy than an exponential and more mass at large $j$.
Since spirals not only shift material around, they also heat it, however, it may be unlikely that spiral activity can be maintained for long enough to achieve this state.
Figure \ref{fig:profiles} shows the $1/R$ profile compared to that of three Sersic profiles with $n=1,2,4$ typical of the range observed for disks ($n=1$) and fast and slow rotators ($n=2-4$).
Many featureless systems appear to get stuck with at most a dynamically hot center surrounded by a warm exponential disk and never make it to the 'Mestel' state.

The addition of dynamically cool material could reinvigorate spiral activity, as considered by, e.g. \cite{sellwood89,sellwood14}.
This could help further compaction and even rebuild back an exponential component in some cases.
This makes the process of AGN feedback, which controls the galaxy's access to gas, incredibly important for whether a pre-Mestel or de Vaucouleur profile is frozen in, finalizing the transition to an early-type galaxy.
In the event that galaxies make it to a Mestel-like state through the help of spiral nexa, the transition to the maximum entropy state envisioned by \citetalias{lynden-bell72} could still be accomplished by spiral groove modes.

\subsection{On the differences between modes and shearing patterns}
The characteristic wave equation for self-gravity dominated spirals (either eq. (\ref{eq:fullchareqn}) or eq. (\ref{eq:long})) makes no explicit distinction between modes or shearing patterns.  However, the nature of the disk response is implicitly controlled by the GDI which determines the speed of both growing and steady long-wave spirals.  When the GDI is either small, such as at inflection point in $\Sigma/\Omega$, or zero, such as in a pure or half-mass (ungrooved) Mestel disk, then the long wave terms vanish and leave only corotation amplification, for which $\omega-m\Omega$ is also zero \citepalias{meidt24}.  The disk response in this case is a wave that shears with the rest of the disk material.  But where the GDI is non-zero, the disk response produces a spiral capable of rotating with a pattern speed that branches off from $\Omega$ (even if the spiral may not be strictly rigidly rotating in the sense it responds with a single speed).  Generally,  exponential disks provide fertile ground for non-zero GDI and mode-like spirals, but there can be situations where the disk conditions force a shearing-like response.  
  
From this perspective, the distinction between 'density waves' and shearing material patterns is an artificial one that is more fundamentally a reflection of conditions in the underlying disk. The spiral multiplicity is also relevant: even where $\omega-m\Omega$ given by the solution to eq. (\ref{eq:long}) is non-zero, it corresponds to an increasingly small difference between $\Omega$ and $\Omega_p$ as $m$ increases.  This, combined with the narrowing of the distance between ILR and OLR as $m$ increases, can make higher $m$ spirals appear to be shearing.  Conversely, low $m$ spirals are more easily recognizeable as genuinely density wave-like.  

The global spiral nexum built at any moment in time out of individual waves that each branch off away from $\Omega$ may also appear to resemble a single, large-scale shearing pattern.  Each branch off $\Omega$ undergoes its strongest amplification near its corotation so that the overall, radially extended appearance can appear dominated by material anchored to $\Omega$ \citep{sellwood14}. 
Careful consideration of spiral phase evolution, combined with spectrogram analyses, are critical for distinguishing between short-lived waves and material patterns in simulations. This evidence can be a powerful tool to calibrate spiral pattern speed measurements \citep[i.e.][]{merrifield06, meidt08a,meidt08b} as a metric for identifying the nature of observed spirals.  

\section{A model for the Quasi-linear Evolution of Individual Spiral Nexa }\label{sec:timedep}
Before calculating the heating and angular momentum exchanges produced by spirals in \S \ref{sec:heating}, we must first complete a basic picture of their growth and decay that can be used to model the spiral's temporal variation.
The motivation is two-fold.  First, with a time-dependent model we can link the characteristic timescale of the spiral's amplitude variation (which is, as yet, unknown) to a given amount of heating or angular momentum change, and vice versa.
Second, and more practically, the characteristic timescale in the model absorbs our lack of knowledge about the actual phase lag between the density and potential (see \S \ref{sec:elements}).
This makes it straightforward to recognize the key ingredients in the disk-spiral feedback loop, which is our ultimate goal.

We begin below by summarizing the properties of spirals at their fastest rate of (instantaneous) growth.
We keep this brief given that our model will be centered on the peak amplitude, where growth is slowest.
(A clearer prediction of the spiral properties at all times would be necessary to build a model for the full evolutionary history of a disk, which is not the goal here.)
Then we use eq. (\ref{eq:long}) to study the longer-timescale evolution of the pattern after a period of disk secular evolution when the disk properties are changed from their initial values.
Thus, we adopt a slightly faster than secular view, in which case eq. (\ref{eq:long}) can also be viewed as constraining $\omega$ as a function of time.  In what follows we will refer to this a quasi-linear evolution.

\subsection{Fastest growing spiral properties for a given set of disk conditions}\label{sec:instantaneous}
In the previous section we identified the conditions in the disk that support spiral growth.
In this section we identify the spiral properties that lead to the fastest growth for a given set of disk conditions.
Part of this investigation was already started in MvdW24, where we used eq. (\ref{eq:fullchareqn}) to investigate growth in the short wave limit, or the emergence of resonant dressing.
The fastest growing of these spirals achieve a tradeoff that maximizes the strength of (self-) gravitational forcing by increasing $k$ while avoiding the stabilizing influence of pressure on small scales.
The result is a  critical pitch angle approximately near 45$\circ$.
Here we will consider specifically long-wave growth that is described by solutions to eq. (\ref{eq:long}) away from corotation (but still far from either the ILR or OLR).
We will focus on the zone $R>R_e$ that allows for growth inside corotation (see section \ref{sec:caverna_architecture}) and thus approximate the GDI term as $\sim -1/R_h$.

In this scenario the characteristic equation becomes
\begin{equation}\label{eq:longnonres}
(\omega-m\Omega)^3\approx2\pi G\Sigma_0\Omega \frac{1}{R_h} \frac{m}{R }\Big[\frac{e^{-i\delta}}{\vert k_{e,long}\vert}-\frac{1}{k_J}\Big].
\end{equation}

Figure \ref{fig:solutions} sketches the solutions to eq. (\ref{eq:longnonres}) in three different regimes defined by the behavior of the potential-density relation.
The black thin and thick lines highlight the solutions that fall inside corotation for at least part of the range in $T_1$.
The thicker black line, in particular, highlights a solution that grows over a large fraction of the plotted range in $T_1$ (discussed more below).
The gray line illustrates the solution for the zone outside corotation.
It corresponds to growth only when $T_1$ is positive (see section \ref{sec:caverna_architecture}).

\subsubsection{The role of the potential-densty phase offset}\label{sec:T1changes}
Through the factor $e^{-i\delta}/\vert k_{e,long}\vert$, the wave's properties -- $k$, $m/R$ and $T_1$ -- determine the rate of spiral growth (or decay) as they establish the strength of self-gravity relative to pressure.
In the specific case of growing spirals inside corotation (i.e. the thick black line in Figure \ref{fig:solutions}), since $T_1$ varies in time as the spiral grows (see the definition of $T_1$ in eq. (\ref{eq:t1def})), the fastest growing (or decaying) solutions to eq. (\ref{eq:fullchareqn}) will also vary in time.  

We will consider three regimes of $T_1$ that correspond to stages in the lifetime of a growing spiral, i.e. one that is undergoing or has undergone growth.
The first (phase 1) represents the wave before $T_1$ has grown substantially.
At this stage, the phase offset $\delta\approx\arctan(kR)^{-1}$, assuming for illustration that $T_0\ll k$.
In the WKB limit $kR\gg 1$ this is negligible, leading to the well-known result that the tight-winding torque density is zero, but it can be positive for long trailing waves.  
Whether $\delta$ is zero or slightly positive, however, eq. (\ref{eq:longnonres}) implies that growth can initiate given the required GDI;
one of the three solutions corresponds to growth inside corotation.
The propensity for growth is due to the donkey effect's ability to quickly introduce and strengthen a negative phase offset, and this exponentially propels further growth.

Indeed, the result of coordinated donkey behavior quickly makes $\delta$ negative, as soon as $T_1<~-1/(2R)$.
From this point, with $\delta<0$, the potential minimum lies securely upstream of the density peak, giving the wave a negative torque density to carry.
In this situation, the net gain of angular momentum experienced by the stars helps keep this wave growing, and even speeds up the growth rate.

The quickened growth associated with a progressively smaller $ T_1$ (moving to the left in the plot) is tied to both an increase in $\delta$ and a decrease in $k_{e,long}$.
The phase of fastest growth (phase 2) is reached when $T_1/R+T_1^2\approx k^2+m^2/R^2$, which makes the phase offset its largest and $k_{e,long}$ its smallest ($k_{e,long}^2\approx \vert 2kT_1\vert $).
Eventually, the wave's amplitude grows so much that $T_1$ dominates the other factors in $k_{e,long}$, changing the dominant radial forcing and reducing the ability of the donkey effect to lead to growth inside corotation.
In this period (phase 3), $T_1^2\gg k^2+m^2/R^2$, and $\delta\approx\arctan(-k/T_1)\rightarrow 0$.
Note that the small positive $\delta$ possible in this case could push growth to outside corotation, as illustrated by the thick black dashed curve in the top panel that moves to the 'outside corotation' region for progressively more negative $T_1$.
In this regime, the solution inside corotation (thin black line), on the other hand, is subject to decay, aided by the increase of $ T_1$ beyond $k_J$, marking an end to the non-resonant structure.

The slowing of growth and the transition to decay is discussed more in Section \ref{sec:disk_evolution}.
Below we first consider the observable spiral properties -- $m$ and/or pitch angle $i_p$ -- associated with fastest growth.

\begin{figure*}[t]
\begin{center}
\begin{tabular}{c}
\includegraphics[width=0.95\linewidth]{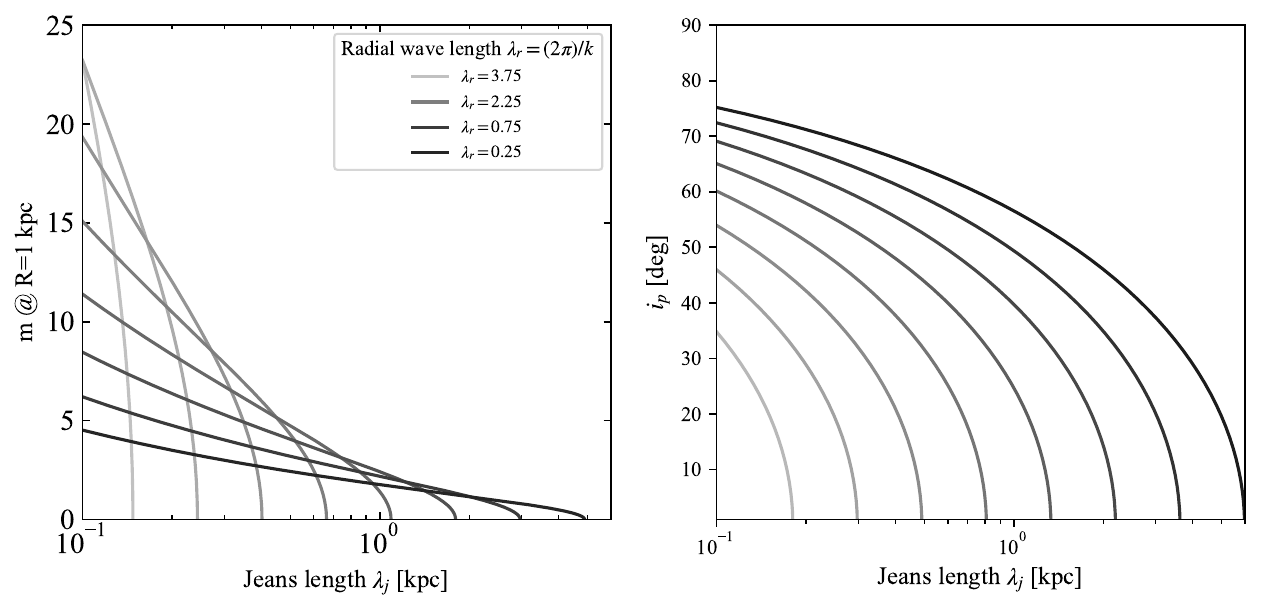}
\end{tabular}
\end{center}
\caption{(Left) An illustration of the spiral arm multiplicity $m$ predicted using eq. (\ref{eq:fastestm}) at a reference radius $R = 1 \text{ kpc}$ plotted as a function of the Jeans length $\lambda_J=(2\pi)/k_J$ (shown in kpc). (Right) The spiral arm pitch angle $i_p$ (in degrees; eq.[\ref{eq:fastestip}]) associated with the $m$ values plotted on the left as a function of $\lambda_J$.  Each line shows a different value for the spiral's radial wavelength $\lambda_r=(2\pi)/k$ (ranging from $0.05$ to $4.0$, from dark to light). Note how dynamically cold disks (small $\lambda_J$) support a wide spectrum of high-$m$, open spirals, whereas warming disks (larger $\lambda_J$) systematically suppress these modes, driving secular consolidation toward low-$m$, tightly wound grand-design patterns.}
\label{fig:pitch}
\end{figure*}

\subsubsection{Early fast-growing wave properties}\label{sec:fastestwaves}
The evolution through phases 1, 2 and 3 carries a wave from a period of fast growth up to its peak in amplitude, where the growth slows and reverses, followed by decay.
For a maturing perturbation to dominate the disk's morphology at its peak, it must presumably have been favored already in phase 1.  With that in mind, in this section we will highlight the initially fastest-growing $m$ and $k$ that we take to reflect the wave-like properties of the spiral over the course of its lifetime.
These fastest growing spiral properties can be identified by seeking the $m$ or $\tan{i_p}$ where the growth rate is maximized.

For the phase 1 growth rate inside corotation we have    
\begin{equation}\label{eq:phase1growthbeta}
\beta_0=\frac{\sqrt{3}}{2}\left(2\pi G\Sigma\Omega \frac{m}{R R_h}\Big[\frac{1}{\vert\sqrt{k^2+\frac{m^2}{R^2}}\vert}-\frac{1}{k_J}\Big]\right)^{1/3}
\end{equation}
in the regime $R>R_h$.
The only real $m^2/R^2$ where this is maximized is 
\begin{equation}
\frac{m^2}{R^2}=k^2\left(\frac{k_j^{2/3}}{k^{2/3}} -1\right)\label{eq:fastestm}
\end{equation}
which is equivalent to 
\begin{equation}
\tan i_p=\left[\left(\frac{k}{k_J}\right)^{-2/3} -1\right]^{1/2}.\label{eq:fastestip}
\end{equation}

Figure \ref{fig:pitch} shows the fastest-growing $m$ (left) and corresponding pitch angle (right) predicted at a radius of 1~kpc using eqs. (\ref{eq:fastestm}) and (\ref{eq:fastestip}), respectively, plotted as a function of $\lambda_J=2\pi/k_J$ at different values of $k$.  The dynamically coldest disks (with small $\lambda_J$) allow for spirals with the greatest $m$ leading to an expansive range of pitch angles.
The lowest pitch angles 5-20$^\circ$ appear when the radial wavelength $2\pi/k$ approaches the Jeans length.
As disks warm and $\lambda_J$ increases, large $m$ and $k$ values are suppressed and this also tends to remove the low end of the range in pitch angles (excepting when $k$ approaches $k_J$.)  

\subsection{The cessation of growth and peak spiral properties}\label{sec:cessation}
\subsubsection{linear evolution}
As alluded to in the previous section, alongside growth, eq. (\ref{eq:fullchareqn}) (or eq. (\ref{eq:longnonres})) also depicts the factors that ultimately bring the growth of individual perturbations in gaseous or stellar disks to a stop.
The principal avenue for halting growth that we consider in this work is related to a weakening and cessation of the donkey effect (inverse Landau damping), which is the process responsible for their amplification.
This happens a result of changes in the dominant forcing term over time, and in particular with the weakening of the self-gravity term in eq. (\ref{eq:fullchareqn}) with respect to the pressure term. (The effects of nonlinear coupling are considered in $\S$ \ref{sec:nonlinearev}.)  
This can be either due to a decrease in the self-gravity or through an increase in the pressure, or both.
The former case occurs as a result of growth changing the potential-density relation (discussed in the previous section) and is also characteristic of shearing patterns.
The latter is the process we have in mind in calculating the secular heating produced by a given non-resonantly growing perturbation.
As described in \citetalias{van-der-wel25}, non-resonant growth signifies that the spiral no longer perfectly transports angular momentum freed at corotation to the ILR with no losses and gets deposited locally in the disk (see also section \ref{sec:heating}). 

We can read from eq. (\ref{eq:phase1growthbeta}) (or eq. [\ref{eq:fullchareqn}]) the properties of the wave at the moment growth turns to decay, which is also the peak amplitude: $\beta_0$ goes to zero when $k_{\rm e,long}e^{i\delta}=k_J=(4\pi G\Sigma)/\sigma^2$. Note that this also immediately suggests that the maximum heating produced by any spiral with $k$ and $m$ can be well approximated by $\sigma^2=4\pi G\Sigma/k_t$ with $k_t\approx k_{\rm e,long}(T_1\approx 0)\approx \sqrt{k^2+m^2/R^2}$.
The peak growth condition $k_t=k_J$ also holds for shearing patterns, in which case it is the weakening of self-gravity rather than heating that inhibits growth.

As described by \citet{sellwood22} and \citet{hamilton24}, donkey behavior can also be halted when perturbations enter a regime in which orbits become non-linearly trapped at corotation \citep[see also][]{sellwood02}.
This trapping saturates mode growth by locally changing the distribution function, removing the feature initially responsible for the spiral's amplification.
This saturation mechanism is indirectly present in eq. (\ref{eq:fullchareqn}) and will be discussed more later in \S \ref{sec:quenching_corotation}.

\subsubsection{nonlinear evolution via mode coupling}\label{sec:nonlinearev}
The evolution sketched so far neglects nonlinear mode coupling \citep{tagger87,sygnet88,masset97}, which starts to become relevant at large density contrasts $\rho_1/\rho_0\gtrsim 0.1$.  This contribution can allow a given mode $m$ to appear for longer than would be predicted from the mode's self-gravity alone (as described below). Even in this case, though, the periodic nature of the beating contribution (in the frame rotating with the mode; see eq.[\ref{eq:s0sqlongNL}]) leaves the mode's time-average evolution tied to self-gravity, thus tying the cessation of mode growth and subsequent evolution to the balance between pressure and self-gravity discussed in $\S$~\ref{sec:cessation}.  

The contribution of non-linear beating to the evolution of a given mode $m$ is encapsulated by the factors in eq. (\ref{eq:s0sqlongNL}) that add non-negligibly to $s_0^2$ when the density contrasts of beating modes $m'$ and $m''$ (satisfying either $m=m'+m''$ or $m=m'-m''$) exceed $\sim$0.1. These modes are influential as long as they also have $k'$ and $m'$ comparable to or less than the mode in question (so that the corresponding potential perturbations are also non-negligible and the term in eq. (\ref{eq:s0sqlongNL}) is relatively large).  This implies that the most important modes $m'$ for powering a given Fourier component $m$ via mode coupling will have $m'<m$, i.e. mode coupling extends Fourier modes from below.  This quality is what allows the mode to linger even after its self-gravity contribution has started to decay, i.e. once $k_t>k_J$ and the growth of any mode higher than $m$ is suppressed.  Again, though, the periodic nature of the beating makes the self-gravity component relevant for the mode's time-average evolution.  

An arguably more important attribute relevant (but implicit) in this work is the ability of mode coupling to seed the next spiral generation \citep[e.g.][]{donghia, chiba25}.  This seeding can occur even when the amplitudes of the beating modes are small.  It could take the form of two evolved (weakened) spiral Fourier components beating against each other or an evolved component  interfering with the response to an independent perturber, for example, as long as the interaction happens faster than its takes for the spiral component to fully evolve (phase-mix) away \citep{chiba25}.  In this case, whether the seed develops into a prominent spiral ultimately rests with the mode's own self-gravity.  This function is implicitly part of how we envision disks are constantly seeded with perturbations, in combination with an abundance of ex situ perturbations (see $\S$ \ref{sec:potential_perts}).  In this scenario, nonlinear mode coupling would coincide with the time of spiral quenching/saturation because the large amplitudes readily seed new perturbations, not because it is responsible for the quenching.  
In section \ref{sec:disk_evolution} we will examine quenching through quasi-linear processes.   
For the purposes of describing the evolution of a single spiral generation, 
an explicit treatment of nonlinear evolution leading to recurrence will be omitted in what follows.

\subsection{A heuristic model for the time-dependence of the spiral potential}
Knowledge of the spiral's peak conditions provides a path towards modelling heating and angular momentum in a way that minimizes the need to specify the spiral's properties at all times.

In what follows, we will employ an approach that breaks the full evolution of the potential into two main regimes, the initial regime and the peak regime (which is then followed by a regime of decay).
The initial regime is when the initial exponential growth rate $\beta_0$ is set and amplifies perturbations with specific properties to prominence.
The peak regime is when any initially fast-growing perturbation reaches its maximum amplitude, i.e. around which the perturbation will exert its largest torque and produce its greatest influence.
For the times between these two regimes, we replace the unknown evolution in $T_1$ (the spatial variation of the potential amplitude) with a model for the time dependence that is (at least partially) the product of that (unknown) spatial variation.
As described below, we stitch the two regimes back together by adopting an exponential that depicts growth at a rate $1/t_s$, with the understanding that $1/t_s\lesssim \beta_0$ given that growth slows on approach to saturation at the peak amplitude.

\subsubsection{The growth phase and saturation phase of spiral nexa}\label{sec:saturationphase}
As our heuristic model for the time evolution of the perturbation amplitude, we adopt an exponential peaking at time $t_{\rm peak}$ with an effective growth rate $\beta_e=1/t_s$, i.e. 
\begin{equation}\label{eq:exp_growth}
\mathcal{F}=\Phi_{1,\rm peak}(R,z)e^{(t-t_{\rm peak})/t_s}
\end{equation}
where $\beta_e<\beta$ (or $t_s\gtrsim 1/\beta$) since $\beta$ is not constant in time but tends to slow towards the peak.

Thus 
\begin{equation}
\int_{-\infty}^{t_{\rm peak}} \Phi_1 dt=t_s\Phi_{1,\rm peak} .
\end{equation}
and 
\begin{equation}
\int_{-\infty}^{t_{\rm peak}} \Phi_1^2 dt= \frac{t_s}{2}\Phi_{1,\rm peak}^2.
\end{equation}

We will rely on this later to express the net change in angular momentum and heating produced by the spiral over the period $-\infty$ to $t_{\rm peak}$ in terms of the observable density at the peak.
Specifically, we use the relationship between the potential and density at the peak to express the total integrated squared potential as
\begin{equation}
\int_{-\infty}^{t_{\rm peak}} \Phi_1^2 dt = \frac{t_s}{2}\left(\frac{2\pi G \Sigma_{1,\rm peak}}{k_J}\right)^2.
\end{equation}

Adding on an exponential decay to eq. (\ref{eq:exp_growth}) after the peak can be used to depict the perturbation's full evolution from growth to saturation and decay, i.e. 
\begin{equation}
\mathcal{F}=
\begin{cases}\Phi_{1,\rm peak}(R,z)e^{(t-t_{\rm peak})/t_s} &t<t_{\rm peak}\\
\Phi_{1,\rm peak}(R,z)e^{-(t-t_{\rm peak})/t_s} &t>t_{\rm peak}.
\end{cases}
\end{equation}

We alternatively opt for the more typical Gaussian time dependence that peaks at $t_{\rm peak}$ with width $t_s$ 
\begin{equation}
\mathcal{F}=\Phi_{1,\rm peak}(R,z)e^{-((t-t_{\rm peak})^2/2)/t_s^2}
\end{equation}
whereby the total integrated squared potential over the full lifetime is
\begin{equation}
\int_{-\infty}^{\infty} \Phi_1^2 dt = \sqrt{\pi} t_s \Phi_{1,\rm peak}^2. 
\label{eq:avgsqpot}
\end{equation}
This is slightly larger than a factor of 2 bigger than the integral up to the peak, although we will treat them as comparable by writing the full integral as $t_s\Phi_{1,\rm peak}^2$.
The growth rate or total lifetime are unknown a priori in both cases, but can be empirically solved using observational constraints (see $\S$ \ref{sec:disk_evolution}).

\subsubsection{Caveats}
Our choice to impose a model for the time evolution makes our calculations comparable to a number of others that seek the response to an adopted external perturbation \citep[e.g.][]{desimone04,binney08,hamilton24c,hamilton26,hamilton25}, though here we focus specifically on the self-consistent response.  
It should be kept in mind, however, that we are not calculating the heating due to self-consistently evolving perturbations, but rather the heating associated with a {\it model} for self-consistently evolving perturbations.
That is, rather than either calculate or model $T_1$ and $T_2$ at all times to infer $\beta$ we adopt a model for the amplitude of the potential that is produced by $\beta$.
This approach is sufficient for building intuition and capturing the basic features of time-integrated heating, which is the main goal here (sections \ref{sec:heating} and \ref{sec:disk_evolution}; as opposed to examining the heating throughout the lifetime of the spiral).  In practice we expect slight deviations from Gaussianity to introduce no greater degree of uncertainty in the predicted heating than associated with empirically constraining either the disk properties or the spiral perturbation properties (amplitude and pitch angle) or the dissipation time.  Still, the model would need to be updated to handle different, significantly non-Gaussian evolution, although this can be accommodated somewhat by the ignorance parameter $\delta_t$ introduced later in $\S$ \ref{sec:timeintegratedtorque}.  

\section{Secular processes in caverna: Dynamical heating and angular momentum transport}\label{sec:heating}

\subsection{Overview}
In this section we explore the angular momentum changes and heating produced by the perturbations described in the previous section.
Here, as there, the focus is on the self-gravity or dressed response and thus any torques associated with non-axisymmetric components of the external potential $\Phi_{ext}$ (satellite galaxies, dark matter halo substructure) are omitted.
It should be kept in mind that the net heating and angular momentum would reflect this additional contribution from the external potential.

\subsection{Torques and Changes in Angular Momentum }\label{sec:torques}

\subsubsection{Instantaneous torque}
Following \citet{goldreich79, binney08} and \citetalias{van-der-wel25}, for any $m$ perturbation we write the instantaneous time rate of change of the angular momentum, or  gravitational torque, in the disk beyond $R$ as 
\begin{eqnarray}
\frac{dL}{dt}=C_g&=&\int_{-\infty}^\infty\int_0^{2\pi}\int_R^\infty \rho\frac{d\Phi}{d\theta} R dR d\theta dz \nonumber\\
&=&\frac{m R}{4G}\Phi_{1,m}^2\frac{k}{\vert k_{e,R}\vert }.\label{eq:totaltorque}
\end{eqnarray}
The second line explicitly accounts for the slow radial variation in the amplitude of the potential (and density) through the factor $k/k_{e,R}$, with $k_{e,R}$ inherited from the potential's vertical distribution.
In the WKB approximation $k_{e,R}=k$ and this factor drops from the expression.
Note that the expression for the torque at the mid-plane calculated without vertical integration retains the factor $k$ (see \S \ref{sec:heatingradialmigration}).

The net gravitational torque is the sum of the torques exerted by each $m$ perturbation.
The total torque is the sum of the gravitational torque and the advective torque, which acts in opposition to the gravitational torque but should be negligible except for the tightest-wrapped spirals (\citealt{binney08}; \citetalias{van-der-wel25}; \citealt{sellwood14a}).
We neglect it here with open spirals in mind, emphasizing that all following calculations apply best outside the tight-winding limit.

\subsubsection{Time-integrated torque}\label{sec:timeintegratedtorque}
To calculate the total angular momentum transported by any single spiral, it is convenient to integrate the gravitational torque produced by each $m$ perturbation either up until the moment it reaches its peak amplitude or over the course of its lifetime.
In the latter case, for example, we integrate the instantaneous torque 
\begin{equation} 
\Delta L_{\rm full} = \int_{-\infty}^{\infty} C_g dt = \int_{-\infty}^{\infty} \left[ \frac{m R}{4G}\Phi_{1}^2\frac{k}{\vert k_{e,R}\vert } \right] dt. 
\end{equation}

In principle this should account for the time-variation in $\vert k_{e,R}\vert $ in eq. (\ref{eq:totaltorque}) as the $T_1$ factor evolves.
Near the moment of fastest growth, for example, $k_{e,R}\ll k$.
In practice, though, we will make the assumption that $k/\vert k_{e,R}\vert$ is approximately unity over the time interval $-\infty$ to $t_{\rm peak}$, as well as over the full lifetime, to arrive at a conservative estimate for the total angular momentum change
\begin{equation}
\Delta L = \delta_t \frac{mR}{4G} \left( \int_{-\infty}^{\infty} \Phi_1^2 dt \right) \approx \delta_t t_s \frac{mR}{4G} \Phi_{1,\rm peak}^2
\label{eq:DL}
\end{equation}
using the integrated squared potential from eq. (\ref{eq:avgsqpot}), and now introducing the order-unity factor $\delta_t$ to represent both the error in our assumed $k/\vert k_{e,R}\vert\approx 1$ and the uncertainty associated with the time dependence of the perturbation's amplitude when slightly different from the assumed Gaussian model.  For evolutionary models that are significantly different from Gaussian, the time integrated calculations in this section should be updated.  Smaller differences can be accommodated through $\delta_t$.  In the case of slightly asymmetric growth and decay differing by a factor $f$, for example, $\delta_t=(1+f)$.  
Numerical simulations that can capture the (non-linear) evolution of various perturbations will be indispensable for calibrating the  ``ignorance parameter" $\delta_t$.  They may also suggest a more suitable alternative to the model adopted in this work.  
In what follows, we use this expression to obtain a picture of the basic features of secular diffusion, although it should be noted that the choice $\delta_t= 1$ may underestimate the full secular diffusion produced by any single spiral.

\subsection{Dynamical Heating }\label{sec:heating2}
The torque exerted at radii beyond $R$ in eq. (\ref{eq:totaltorque}) leads to an increase in the angular momentum of orbiting material.
To support this pattern of angular momentum increase, material at radii inside $R$ must lose angular momentum.
This loss of angular momentum inside $R$ liberates the energy in circular motion for non-circular motion \citep{lynden-bell72, binney08, van-der-wel25}.

The change in energy density due to interaction with the spiral can be calculated as
\begin{equation}
\frac{dE}{dt}=\int_{-\infty}^\infty\int_0^{2\pi}\int_R^\infty \rho\frac{\partial \Phi}{\partial t}RdRd\phi dz.
\end{equation}
In terms of the angular momentum density $\Delta l$ gained in some time $t$, this yields $\Delta E(t)=-(\Omega-\Omega_p)\Delta l(t)$ for steady, rigidly rotating patterns and $\Delta E(t)=-(\Omega-\Omega_p+\eta)\Delta l(t)$ for patterns with continuous amplitude and/or shape changes represented by $\eta$. For purely exponential growth at rate $\beta$, for example, the factor $\eta= \beta T_1/(k_{e,R}m)$, while for our assumed Gaussian time dependence $\eta= (t_{peak}-t)/t_s^2 T_1/(k_{e,R}m)$.  Models in which the variation in $T_1$ and $k_{e,R}$ are not symmetric about the peak amplitude retain a factor $\eta_{\rm eff}$ proportional to $\eta$ in the expression for heating.  This can be used to capture orbit trapping and heating at corotation, as well as heating by shearing material patterns, for example.  An estimate of $\eta_{\rm eff}$ and a discussion of its role is postponed until later in $\S$ \ref{sec:nonresHeat1}.  

Written as a change in velocity dispersion, the change in energy density for a change in angular momentum implies 
\begin{equation}
\frac{d\sigma^2}{d t}=\frac{-(\Omega-\Omega_p+\eta)}{2\pi \Sigma R} \frac{d}{dR}\left(\frac{ d L}{d t}\right).
\end{equation} 
Integrating over the full lifetime of the spiral, we find the total heating $\Delta\sigma^2$
\begin{equation}\label{eq:rate0potential}
\Delta\sigma^2 = \frac{-\mathcal{W}_s}{2\pi \Sigma R} \frac{d}{dR} (\Delta L), 
\end{equation} 
which is directly proportional to the total angular momentum exchanged.  Here we let
\begin{equation}
\mathcal{W}_s=\begin{cases}(\Omega-\Omega)&\text{steady spiral modes}\\
(\Omega-\Omega+\eta_{\rm eff})&\text{non-steady spiral waves}\\
\eta_{\rm eff}&\text{shearing material spirals}
\end{cases}
\label{eq:wscases}
\end{equation}
using $\mathcal{W}_s$ generically to denote the effective rate at which the spiral changes form.  

Substituting our expression for $\Delta L$ (eq.~\ref{eq:DL}) and taking the radial derivative, this yields
\begin{equation}\label{eq:rate2potential}
\Delta\sigma^2 = \frac{\mathcal{W}_s \delta_t t_s m \Phi_{1,\rm peak}^2\mathcal{F}_{\rm NR}}{8\pi  G\Sigma R} 
\end{equation}
in terms of 
\begin{equation}
\mathcal{F}_{\rm NR}=-\left(1+\frac{d\ln \Phi_{1,\rm peak}^2}{d\ln R}\right)
\end{equation}
where $d\ln \Phi_{1,\rm peak}^2/d\ln R=2T_1R$.
The factor $\mathcal{F}_{\rm NR}$ reflects the potential's amplitude at its peak, which is the product of the differential growth experienced over time.
Note that this corresponds to net heating as long as $2T_1R<-1$, which is identical to the criterion for a negative potential-density phase lag and negative torque density inside corotation (see $\S$ \ref{sec:wkb}; see also \citealt{zhang96,dehnen25}).  

The heating $\Delta\sigma^2_{\rm peak}$ produced in the time interval leading up to the peak is half of this value, i.e.
\begin{equation}
\Delta\sigma^2_{\rm peak} = \frac{1}{2}\Delta\sigma^2 \label{eq:halfheat}
\end{equation}
given that the integrated squared potential up to the peak is half of the full integral.

\subsubsection{The spatial location of growth, heating and decay}
In practice, the factor $\mathcal{F}_{\rm NR}$ delimits where in the disk the largest secular changes occur, determined by whether the growth is resonant or non-resonant and whether the pattern is steady (in which case $ d C_g/dR=0$; \citealt{goldreich65, binney08}) or non-adiabatic (with $ d C_g/dR\neq0$; \citetalias{van-der-wel25}).
In the case of resonant growth predicted for adiabatic perturbations, the self-gravity potential perturbation is tightly centered on corotation making $\vert d\ln \Phi_{1,\rm peak}^2/d\ln R\vert \gg 1$ in the immediate vicinity and bringing $d C_g/dR$ to zero elsewhere.
Since $\mathcal{W}_s = 0$ at corotation for steady rigid patterns, there is no heating there, nor at any other radius where $dC_g/dR = 0$ (or $\mathcal{F}_{\rm NR} = 0$).
(In the tight-winding limit, changes to $T_1$ are negligible and $\mathcal{F}_{\rm NR}$ never substantially increases.) 
 The only other locations with large $\vert d\ln \Phi_{1,\rm peak}^2/d\ln R\vert $ are where the perturbation reaches its physical limit and gets absorbed.
This makes the ILR a major source of heating for this type of perturbation (\citealt{toomre69}, \citetalias{lynden-bell72}, \citealt{sellwood14a}).
Given the absence of corotation heating, the growth for such patterns is stopped through different processes, including a flattening of the distribution function through non-linear orbit trapping (\citealt{sellwood22, hamilton24}; see section \ref{sec:disk_evolution}).

For open, non-adiabatic, non-resonantly growing patterns or shearing material patterns, both of which are favored when $kR\sim 1$, a zone with non-zero $d C_g/dR$ widens around corotation, allowing the factor $\mathcal{F}_{\rm NR}$ to become non-negligible (and once again reaching a maximum at the spatial limits of the spiral).
Now, spiral torques can lead to exchanges of angular momentum and heating away from resonances (\citetalias{van-der-wel25} and see section \ref{sec:nonresHeat1}) as well as at corotation (see section \ref{sec:nonresHeat1}).  
Given their potential for producing widespread secular changes, non-steady, non-adiabatic spirals \citepalias{van-der-wel25} that grow to prominence away from corotation (and have $d C_g/dR\neq0$ by definition) are of greatest interest in this work.
In \S \ref{sec:instantaneous} we showed that spirals grow fastest when $T_1\approx -\sqrt{k^2+m^2/R^2}$, suggesting that the spiral's peak amplitude (and slowed growth) may coincide with $\vert T_1\vert<k_t$.
We will conservatively require $\vert T_1 R\vert< m$ making $\mathcal{F}_{\rm NR}$ a quantity approaching unity, which is also consistent with the choice $\delta_t\approx 1$.

Generally, any perturbations with $\vert T_1 R\vert$ approaching or exceeding $m$ are deemed to lack the wave-like quality relevant to the present study.
Observations of spirals in the stellar disks of nearby galaxies (and now also at higher redshift) do indeed suggest that spirals are broad in extent (rather than limited to a narrow corotation zone) and exhibit only mild changes in amplitude.
This suggests that $\mathcal{F}_{\rm NR}$ is of order unity at times around the peak amplitude of a given perturbation (although it could be much larger, for example at the edges of the perturbation).

\subsubsection{Non-resonant and resonant heating in terms of the perturbed density I}\label{sec:nonresHeat1}
In this section we wish to use eq. (\ref{eq:rate2potential}) to get a sense of the magnitude and characteristics of the heating produced by a spiral perturbation with a given set of properties.
We begin by using the potential density-relation in eq. (\ref{eq:pot_dens_rel}) to rewrite the heating over the spiral lifetime in eq. (\ref{eq:rate2potential}) in terms of the observable density finding   

\begin{equation}\label{eq:rate2density}
\Delta\sigma^2\approx t_s\mathcal{W}_s\delta_t\mathcal{F}_{\rm NR} \frac{m}{ 4R} \frac{2\pi  G\Sigma}{k_{e,\rm peak}^2}\left(\frac{\Sigma_{a,\rm peak}}{\Sigma }\right)^2
\end{equation}
where $k_{e, \rm peak}$ is given by $k_{e, \rm long}$ in eq. (\ref{eq:kelong}) at the moment the spiral reaches its peak amplitude and the factor $\mathcal{F}_{\rm NR}$ allows that this is non-zero away from corotation.  

Before considering how the spiral strength, multiplicity and pitch angle impact heating (below, section \ref{sec:nonresHeat2}), it is worth first noting the role of the spiral lifetime $t_s$ and the factor $\mathcal{W}_s$.  
For non-steady spirals with our adopted Gaussian time-dependent amplitude the factor $\mathcal{W}_s$ includes a term 
\begin{equation}
\eta_{\rm eff}=\frac{1}{t_s}\frac{\int_{-\infty}^{\infty} \frac{(t_{peak}-t)}{t_s^2}\frac{T_1}{k_{e,R}m}\frac{d}{dR}\frac{d L}{dt}dt}{\int_{-\infty}^{\infty}\frac{d}{dR}\frac{d L}{dt}dt}\label{eq:eta}
\end{equation}
(see section \ref{sec:heating2}).  
Both $T_1$ and $k_{e,\rm R}$ also vary with time, and it is the asymmetric quality of the variation about the spiral's peak amplitude, in addition to the finite lifetime, that allows $\eta_{\rm eff}$ to be non-zero.  (With fixed $T_1$ and $k_{e,R}$, eq. [\ref{eq:eta}] integrates to zero.)  Wave perturbations that amplify most strongly start with $\vert T_1/k_{e,R}\vert\ll 1$, and this factor reaches a maximum at peak growth.  Ultimately, though, $\vert T_1/k_{e,R}\vert\rightarrow 1$.  This will tend to keep the overall contribution to $\eta$ negligible on the growth side of the peak but non-zero on the decay side.   
Taking $T_1/k_{e,R}\vert\approx 1$ only from the peak onwards, eq. (\ref{eq:eta}) yields a minimum $\eta_{\rm eff}\approx 1/(2m t_s)$.  

This suggests that the contribution of $\eta_{\rm eff}$ to $\mathcal{W}_s$ is only substantial when $t_s\lesssim(\Omega-\Omega_p)^{-1}$ and becomes increasingly negligible for spirals with the longest lifetimes, in which case the heating expression behaves exactly as in the case of steady waves.  Still, for non-steady, short-lived spirals, $\eta_{\rm eff}\neq0$ opens the possibility of heating at corotation or for shearing material spirals with $(\Omega-\Omega_p)=0$.  We will proceed assuming that the spiral lifetime is finite but long enough that $\eta_{\rm eff}$ sets a floor to the heating that is otherwise raised above this level away from corotation.  
 Near corotation, we expect secular changes to be dominated by angular momentum exchanges between the wave and the stars (\citealt{sellwood02}; see \S \ref{sec:quenching_corotation} later).

 \subsubsection{Non-resonant and resonant heating in terms of the perturbed density II}\label{sec:nonresHeat2}

The net heating implied by eq. (\ref{eq:rate2density}), calculated as the sum of the heating due to all structures (over $m$), will be dominated by perturbations with the largest $\Sigma_{a,\rm peak}$  and $m$ (or $\tan i_p$).
With this in mind, eq. (\ref{eq:rate2density}) provides a way to observationally estimate the heating produced by a given spiral, taking the spiral's arm number as the dominant $m$ and assuming that it is observed at or near its peak.  The measured pitch angle and radial profile of the spiral's amplitude could then be used to place constraints on $k_{e,\rm peak}^2$.

More general inferences about the degree of heating can be made based on the expected properties at the peak.
For example, if the heating is very effective, the peak could be reached already when $T_1\approx 0$ in which case $k_{e,\rm peak}^2$ would be comparable to $k^2+m^2/R^2$.
It may instead be expected likely to fall between $k_{e,\rm peak}^2=kT_1$, the value at fastest growth (when $T_1/R+T_1^2=k^2+m^2/R^2$), and $k_{e,\rm peak}^2\approx T_1^2$, the value in the case that (at some presumably later period in time) $\vert T_1\vert\gg k$.

These three possibilities provide us with the following upper and lower bounds: 
\begin{eqnarray}
\Delta\sigma^2&\approx& t_s\mathcal{W}_s\delta_t \pi  G\Sigma R\tan{i_p}\left(\frac{\Sigma_{a,\rm peak}}{\Sigma }\right)^2 \\
&\times& 
\begin{cases}
\frac{1}{2kR} & \text{with } k_{e,\rm peak}^2\approx k^2~\text{and}~\mathcal{F}_{\rm NR}\approx 1 \\
1 & \text{with } k_{e,\rm peak}^2\approx k T_1 ~\text{and}~\mathcal{F}_{\rm NR}=\vert 2T_1R\vert \\
\frac{k}{T_1} & \text{with } k_{e,\rm peak}^2\approx T_1^2 ~\text{and}~\mathcal{F}_{\rm NR}=\vert 2T_1R\vert 
\end{cases}\nonumber
\end{eqnarray}
using that $\mathcal{F}_{\rm NR}\approx \vert 2T_1R\vert $ more appropriately when $\vert T_1\vert\sim k$ in the latter two cases.

The heating in these estimates is clearly most effective for open spirals and those with slowly varying amplitudes (see also \citealt{hamilton24c}).
We will take the upper bound in the second of these cases to provide a quantification of the heating, together with a few additional assumptions.
First, we arbitrarily adopt a short spiral lifetime $t_s\sim 1/\mathcal{W}_s$, so that our estimate becomes a lower upper bound on the heating.
We also take $\pi  G\Sigma R\approx V_c^2/2$.  This suggests   
\begin{equation}
\frac{\Delta\sigma^2}{V_c^2} \approx \frac{\tan i_p}{2} \left(\frac{\Sigma_{a,\rm peak}}{\Sigma}\right)^2.
\end{equation}

Selecting $i_p=20^\circ$ and $\Sigma_{a,\rm peak}/\Sigma \approx 0.1$ as representative of the $m=2$ spirals in typical disk galaxies (e.g. \citealt{yu20}; see also \citetalias{van-der-wel25}) this yields $\Delta\sigma/V_c\sim 0.04$ after one non-resonant spiral generation.

The heating estimate presented here can be made quantitative for a given spiral with knowledge of the spiral lifetime and the value of $k_{e,\rm peak}$, which in turn requires a constraint on the value of $T_1$ at the spiral's peak (i.e. measured directly from the density distribution).  
To date, few spiral lifetime estimates exist, and few, if any, empirical measurements of the amplitude variations of either the density or potential have been made in a way that provides a typical picture of spirals.  
This would be an ideal area for concerted effort in the near future.

\section{The evolution of disks and their structures}\label{sec:disk_evolution}

In this section we will examine how the changes in angular momentum and heating produced in a disk by any one spiral nexum set a natural limit to the lifetime of different components of that spiral.
Stellar dynamical theory and simulations have already demonstrated how spirals saturate at corotation or in grooves as a result of orbit trapping, phase mixing and the subsequent reshaping of the density distribution through radial migration \citep[i.e][]{sellwood22,hamilton24}.
We will add to that picture three additional processes that amount to a quenching of spiral growth.
The first of these is related to the heating produced by the non-resonant part of spirals, which raises the velocity dispersion, overwhelming the gravitational force and locally shutting down growth.
The second is a weakening of self-gravity due to growth's influence on the potential-density relation.
The third is similar to groove saturation and involves the redistribution of mass that removes the caverna features responsible for stimulating spiral activity.

Together, the lifetimes calculated in this section build a framework for comparing our hydrodynamical approach to stellar dynamical calculations and numerical simulations.
It is worth noting that differences may be expected between the hydrodynamical view of spiral decay and the process of saturation and decay through inherently kinetic phenomena like phase mixing, distribution function scattering and fine-grained resonant 
orbit trapping.  These microscopic processes are smoothed over when replacing the full distribution function with a 
macroscopic density and pressure and occur alongside the macrophysical quenching mechanisms discussed below.  
With this in mind, we note that our view of how heating limits the spiral lifetime may underestimate the true lifetime.  When phase mixing is responsible for spiral decay (as opposed to an increase in velocity dispersion), the structural remnants of the spiral may persist a few orbital periods longer as they wrap, extending the lifetime and allowing their influence to continue \citep[e.g][]{chiba25}.  

\subsection{(Quasi-) Linear Quenching due to Heating}\label{sec:quenching_heating}
The spiral lifetime is a key factor determining the amount of heating produced by a spiral (see eq. (\ref{eq:rate2density}), previous section).
At the same time, the heating itself amounts to a local weakening of self-gravity\footnote{In our hydrodynamical approach, the weakening of self-gravity is through an increase in the effective pressure in the gas.} and thus becomes a principal factor limiting the lifetime.
In this section we access that role by explicitly aligning the peak with the moment $k_{e,long}=k_J$, in terms of $k_J=2\pi G\Sigma/\sigma^2$.
Substituting $k_{e,long}=k_J$ in eq. (\ref{eq:halfheat}), we have 
\begin{equation}
\frac{\sigma_{\rm peak}^2-\sigma_0^2}{t_s}\approx \mathcal{W}_s \frac{\sigma_{\rm peak}^4}{2\pi  G\Sigma} \frac{m}{8 R}\left(\frac{\Sigma_1}{\Sigma }\right)^2.
\end{equation}

Note that $k_{e,long}$ can become equal to $k_J$ through both a decrease in $k_J$ and an increase in $k_{e,long}$ as growth increases $T_1$.
This quadratic equation for $\sigma_{\rm peak}^2$ has as its solution
\begin{equation}
\sigma_{\rm peak}^2=\frac{\sigma_{\rm heat}^2}{2} \left(1\pm \sqrt{1- 4 \frac{\sigma_0^2}{\sigma_{\rm heat}^2}}\right)\label{eq:sigsoln}
\end{equation}
in terms of 
\begin{equation}
\sigma_{\rm heat}^2=\frac{16 \pi  G\Sigma_0}{ \mathcal{W}_s t_s}  \frac{R}{m}\left(\frac{\Sigma_0}{\Sigma_1 }\right)^2.
\end{equation}

Eq. (\ref{eq:sigsoln}) yields real, physical solutions as long as $\sigma_0^2/\sigma_{\rm heat}^2<\frac{1}{4}$.
Taking the limit $\sigma_0^2\ll  \sigma_{\rm heat}^2$, for example, we have either the trivial solution $\sigma^2=\sigma_0^2$ or the maximum heating solution $\sigma^2=\sigma_{\rm heat}^2$ with $\sigma_{\rm heat}^2\gg \sigma_0^2$.
As $\sigma_0^2$ approaches $\sigma_{\rm heat}^2/4$ we obtain the more conservative solution $\sigma_{\rm peak}^2\approx 2\sigma_0^2=\sigma_{\rm heat}^2/2$, or 
\begin{eqnarray}\label{eq:sigsoln_diff}
\sigma_{\rm peak}^2-\sigma_0^2&=&\frac{\sigma_{\rm heat}^2}{4} \nonumber\\
&=&\frac{2 \pi  G\Sigma_0}{ \mathcal{W}_s t_s}  \frac{2 R}{m}\left(\frac{\Sigma_0}{\Sigma_1 }\right)^2.
\end{eqnarray}

Now we have an expression that encapsulates the intrinsic link between heating and lifetime while also illustrating which spiral properties lead to the most heating.
We can infer, for instance, that the shortest-lived spirals are those that have led to rapid heating.
The strongest spirals moreover produce less integrated heating because the heat they generate quickly limits their lifetime.
Similarly, a large $m$ reduces the heating output because the large changes in angular momentum lead to such efficient heating that the spiral's lifetime is reduced.
Another way of illustrating this link between spiral properties and lifetime is to transform eq. (\ref{eq:sigsoln_diff}) into an expression for the quenching amplitude, i.e. 
\begin{equation}\label{eq:saturation}
\frac{\Sigma_{1,\rm peak}}{\Sigma}=\left(\frac{1}{\mathcal{W}_s t_s}\frac{2R}{m(k_J^{-1}-k_{J,0}^{-1}) }\right)^{1/2} 
\end{equation}
where $k_{J,0}$ and $k_J$ are the inverse Jeans length at the start of the heating and at peak heating, respectively.
This expression alternatively gives us a measure of the spiral lifetime,  
\begin{equation}\label{eq:timescale}
t_s=\frac{1}{\mathcal{W}_s}\frac{2R}{m(k_J^{-1}-k_{J,0}^{-1}) }\left(\frac{\Sigma}{\Sigma_{1,\rm peak}}\right)^2.
\end{equation}
The lifetime is longest near corotation where $\mathcal{W}_s$ reaches its minimum (eq. \ref{eq:wscases}).  

Eq. (\ref{eq:timescale}) highlights that the weaker perturbations are able to heat the disk for longer periods of time than stronger perturbations.
It also shows that the lowest $m$ heat by a greater degree, giving them the most influence.
This dependence on $m$ has interesting implications for the efficiency of heating over time.
Consider that early, dynamically cooler disks can host spirals with higher $m$, but as the disk warms, spiral growth shifts to lower $m$.
This will tend to make later spirals longer lived and engaged in heating over longer periods.   

The lifetime in eq. (\ref{eq:timescale}) is also clearly linked to the present (and past) dynamical state of the disk.
The shortest lifetimes are associated with the largest changes in velocity dispersion, i.e. $1/k_J \gg 1/k_{J,0}$.
The lifetime in this case be inferred from how different $m/R$ is from $k_J$.
Dissipation, in the form of the addition of dynamically cool young stars, on the other hand, can keep the difference in $k_J$ and $k_{J,0}$ small, providing a path to lengthen the lifetime (see Sellwood \& Carlberg).  (Cooling can also be expected to impact the fastest-growing $m$.) 
The overall effect is to lengthen both the individual spiral episodes and the duration of the disk's full period of spiral activity.
The shortest of the lifetimes predicted by eq. (\ref{eq:timescale}) can be estimated assuming that heating is efficient ($k_J\ll k_{J,0}$), in which case  
\begin{equation}
t_s\sim \frac{t_{\rm cross}}{2\pi}\left(\frac{k_JR}{m}\left(\frac{\Sigma_0}{\Sigma_{peak}}\right)^2-\frac{1}{2m}\right)\label{eq:simpleheat}\\
\end{equation}
in terms of the spiral crossing time $t_{\rm cross}={2\pi}/(\Omega-\Omega_p)$ and replacing $\mathcal{W}_s$ with eq. (\ref{eq:wscases}) using the expression for $\eta_{\rm eff}$ that corresponds to a floor level of heating associated with the spiral's non-steadiness introduced in section \ref{sec:nonresHeat1}. Here again the tighter, lower $m$ spirals are implied to live the longest.

Note that the first term in parentheses $k_JR (\Sigma_0/\Sigma_{peak})^2$ will mostly greatly exceed $1/2$.  This implies that (except in already dynamically hot systems), corotation heating poses little limitation to the lifetime at corotation, especially compared to the faster mechanisms considered in the following next sections. Generally, the longer $t_{\rm cross}$, the longer the lifetime.  

A rough estimate for the non-resonant lifetime can be obtained by letting the heating produce an early peak so that $T_1\approx 0$ (weak amplitude radial variation) and $k$ and $m/R\sim k_J$, in which case eq. (\ref{eq:simpleheat}) simplifies further to $t_s\sim t_{\rm cross}/(2\pi)\left(\Sigma_0/\Sigma_{peak}\right)^2$.
Using that observed and simulated density contrasts are in the range 0.1-0.3 \citep[e.g][]{elmegreen11,yu20,fujii11,donghia,sellwood22}, then far from corotation where $t_{\rm cross}\sim t_{\rm orb}$, the lifetime will be as short as 2-16 orbital periods (or about 1-10 orbital periods at the effective radius).  
The shortest lifetimes are comparable to the timescales for shearing material patterns to emerge and decay.   With $T_{\rm orb}\sim 200$~Myr at the effective radius (such as for massive galaxies in the local universe \citepalias{van-der-wel25}), this corresponds to anywhere between several hundred Myr and a few Gyr.  As discussed later in $\S$ \ref{sec:discussion}, cosmological zoom-in simulations (e.g. Auriga and FIRE-2; \citealt{grand26,quinn25}) produce spirals that evolve on similarly short timescales.   

This estimate conversely suggests that transient spirals lasting for a few orbital times undergo quenching almost immediately upon entering the non-linear regime.  This again is consistent with the predominantly linear evolution found in N-body simulations, with amplitudes rising to maximum values $\Sigma_{peak}/\Sigma_0\sim0.1-0.3$ \citep{fujii11,donghia,sellwood22}. 

\subsection{(Quasi-) Linear Quenching due to weakening self-gravity (and heating)}\label{sec:bottomsdream}
Given that the heating becomes less effective close to corotation, the quenching lifetime due to heating invariably increases moving toward corotation (see eq.[\ref{eq:simpleheat}] previous section).
Here, though, spirals can take different paths to quenching \citep[e.g.][]{sellwood22}.
One of these paths is through a saturation process through orbit trapping, as examined by \cite{hamilton24}.
Another, considered in this section, stems from the change in the potential-density relation as the spiral's amplitude grows.
This is the `Bottom's Dream' mechanism originally discussed in the short-wave limit by \citetalias{meidt24}.
Essentially, as the spiral grows, the gradient of the spiral's envelope -- the factor $\vert T_1\vert $ -- also increases.
Following the period in which growth can be sped up (see \S\ref{sec:T1changes}), eventually $\vert T_1\vert $ increases enough that not only the dominant component of the radial gravitational force changes, but, more importantly for the azimuthal direction and the long-wave terms in eq. (\ref{eq:fullchareqn}), the perturbation's self-gravity is directly weakened. This disengages donkey behavior and slows and eventually halts the spiral's growth.
In our hydrodynamical approximation, this process is captured by the decrease in the self-gravity term below the pressure term in the characteristic equation.

This also makes it linked to the process of heating described in the previous section (and indeed, the weakening of self-gravity contributes to the timescale estimated there).
The timescale derived in this section should thus also be thought of as heating timescale, though with an alternative emphasis on the variation in the potential-density relation over time.

Given that the idea here is mainly to capture the importance of $T_1$, we will assume that this mode of saturation becomes important once $T_1^2\gg k^2+m^2/R^2$, which also coincides with a significant reduction in the potential-density phase offset.
In this case, `Bottom's Dream' quenching occurs when $T_1^2\approx k_j^2$.
The moment in time $t_{s,BD}$ associated with this quenching can be identified using that 
\begin{equation}
T_1(t_{s,BD})^2=\left(T_{1,0}+\int_0^{t_{s,BD}} \frac{d\beta}{dR} dt'\right )^2.
\end{equation}

Several additional assumptions allow us to estimate $t_{s,BD}$.  First, we assume that the initial value $T_{1,0}$ is negligible compared to the initial growth.
Second, for the radial gradient of the growth rate $\beta$ we will take $d \beta/dR=-\beta/R$ (given the radial variation in $\Omega$ and $m/R$ in $\beta$ in eq. (\ref{eq:phase1growthbeta})) and approximate the integral of $d\beta/dR$ over time as $-\beta_0/R t$, where $\beta_0$ is the initial growth rate.
This is an overestimation of the true integral given that $\beta$ decreases from $\beta_0$ over time (reaching zero at $t_{s,BD}$), but it allows us to write the following lower bound on $t_{s,BD}$, 
\begin{equation}
t_{s,BD}=\frac{k_j R}{\beta_0}.
\end{equation}

Now we substitute in an estimate of the original growth rate, which we assume roughly coincides with $T_1=0$ and reflects a dominance of self-gravity over pressure.
In this case the growth rate in eq. (\ref{eq:phase1growthbeta}) becomes 
\begin{equation}
\beta_0\approx \Omega\left(\frac{m 2z_0}{Q_M R R_e}\frac{1}{\sqrt{k^2+\frac{m^2}{R^2}}}\right)^{1/3}
\end{equation}
where $Q_M=\kappa^2/(4\pi G\rho_0)\approx 2 Q$ \citep{meidt22} in terms of the Toomre $Q=\sigma\kappa/(\pi G\Sigma_0)$ and using that $\Sigma_0=\rho_0 2z_0$ where $z_0$ is the vertical scale height of the disk.
This implies a quenching timescale
\begin{equation}\label{eq:tbd}
t_{s,BD}\gtrsim \frac{k_j R}{\Omega}\left(Q_M \cot{i_p} (1+\tan i_p^2 )^{1/2}\frac{R_h}{2 z_0}\right)^{1/3}
\end{equation}

Observations of nearby stellar disks suggest $z_0/R_h\sim 0.4$ \citep{yu26}.
Adopting $Q_M\sim 3-4$ and setting $i_p\sim 20^\circ$, eq. (\ref{eq:tbd}) yields a timescale $t_{s,BD}$ that is just two orbital periods (at fastest), comparable to the timescale in eq. (\ref{eq:timescale}), though indeed shorter in the zone around resonance.  Thus, even where heating becomes ineffective, spirals face another strong, self-imposed challenge to their longevity.
To avoid this limit and live for longer, spirals would need to undergo less strong differential growth so that $T_1$ is kept small over time.

The $\cot i_p$ factor in eq. (\ref{eq:tbd}) makes the lifetime sensitive to the spiral multiplicity.  Similar to quenching due to non-resonant heating, `Bottom's Dream' lifetimes are longest for tighter-winding (lower $m$) spirals.  

\subsection{The (Quasi-) Linear Saturation and quenching at corotation}\label{sec:quenching_corotation} 
Aside from the heating and weakening of self-gravity that together limit spiral lifetimes, there are two additional processes that become influential especially near corotation.
The first is active when the spiral's corotation is situated in a groove in the disk's distribution function.
The second applies to a generic corotation in an exponential disk.

In the case of growth prompted by a groove in a Mestel disk, for example, \cite{hamilton24} argues that saturation occurs when the trapping time $t_{\rm trap}=2\pi/\beta$, resulting in saturation to amplitude $\Sigma_a/\Sigma_0\approx 2(\beta/\Omega_p)^2 \cot{i_p}$ once the angular momentum changes reshape and flatten out the initial groove.
This corresponds to a saturation timescale $t_{\rm sat}\sim t_{\rm orb} (\cot{i_p} \Sigma_0/\Sigma_a/2)^{1/2} $.
For our canonical $\Sigma_a/\Sigma_0\sim 0.1-0.3$ and $i_p=20^\circ$ this saturation occurs after 2-4 orbits, comparable to $t_{\rm BD}$ and shorter than $t_{\rm heat}$ near corotation.

Wave growth (or dressing) at corotation as the result of some other (external) excitation, rather than a groove, undergoes a similar self-inflicted damping.
Changes in angular momentum locally reshape the disk and, once this reshaping leads to a Mestel-like distribution, spiral amplification (dressing) shuts off.
This process can last considerably longer than it takes to fill in a groove, however, especially at large galactocentric radius where the exponential is increasingly less Mestel-like to start.

To model the reshaping process, we use that torques and angular momentum exchanges between the disk and the wave after some time $t_s$ yields a change in the disk's surface density by an amount 
\begin{equation}
\Delta \Sigma=\frac{t_s}{2\pi R V_c}\frac{\partial}{\partial R}\frac{m\Phi_{\rm peak}^2}{4\pi G}\mathcal{F}_{NR}
\end{equation}
(see \citetalias{van-der-wel25}).
For the corotation dressing, we assume that $\Phi_{peak}$ is a Gaussian centered on the corotation radius $R_{CR}$ with a width $\Delta R$.
In this case, 
 \begin{equation}\label{eq:deltasigma1}
\Delta \Sigma\approx \frac{t_s}{2\pi R V_c}\frac{m}{4\pi G}\Phi_{\rm peak,0}^2 \frac{2R_{CR}}{\Delta R}
\end{equation}
where $\Phi_{\rm peak,0}^2$ is the temporal peak of the potential at the Gaussian's center at $R_{CR}$.

Now we calculate the time required for the radial gradient of total surface density profile to be changed from $\propto -\Sigma/R_e$ to $\propto -\Sigma/R$.
The matching  condition $(1/\Sigma_{tot})d\Sigma_{tot}/dR=-1/R$ implies that, across the zone of width $\Delta R$, we must have
\begin{equation}\label{eq:deltasigma2}
\Delta \Sigma=\Sigma\left(\frac{R}{R_e}-1\right)\left(\frac{\Omega}{\beta}+1\right)^{-1}
\end{equation}
using that the width of the corotation zone $\Delta R=(R\Delta\Omega)/\Omega=R\beta/\Omega$ (Hamilton 2025).
Equating eq. (\ref{eq:deltasigma1}) in the limit $\beta/\Omega\ll 1$ with eq. (\ref{eq:deltasigma2}) yields the timescale $t_s$, 
\begin{equation}
t_s=t_{orb}\frac{V_c^2}{2\pi G\Sigma R}2m \cot^2 i_p \left(\frac{\beta}{\Omega_p}\right)^{3/2} \left(\frac{R}{R_e}-1\right)\left(\frac{\Sigma}{\Sigma_1}\right)^2
\end{equation}
where for simplicity in this estimate we have taken $\Phi_{\rm peak,0}=2\pi G\Sigma_1/k$ in the WKB approximation.

In the Mestel-like region near $R\sim R_e$ in a fast-rising rotation curve, this timescale can be quite short (a few orbital periods).
However, the timescale lengthens considerably at larger $R$, where it is roughly an order of magnitude longer than the other timescales considered above.
Reshaping an exponential into a Mestel-like $R^{-1}$ profile takes considerably longer the further out the spiral is positioned in the exponential, where it is increasingly less Mestel-like.
The implication is that outer resonant dressing can be longer-lived than resonant dressing present at smaller radii.
The latter features instead can evolve almost as rapidly as the non-resonant component of spiral nexa.

\subsection{Discussion}\label{sec:discussion}
\subsubsection{The lifetimes of resonant and non-resonant Spiral activity}
Our calculation of lifetimes in this section establish a picture of spirals as intrinsically multi-component, self-limiting structures.
The changes to disk properties they produce, together with the evolution of their own properties over time, necessarily limits the length of time any single spiral can last.
Remarkably, the non-resonant and resonant components of spirals evolve on similar (though not identical) timescales, growing and decaying both within several orbital periods at their fastest.
This suggests that the non-resonant component may be as important for the disk's morphology and dynamics as the resonant component over the course of the spiral's lifetime.

In the inner disk, subtle differences in lifetime along a spiral (given changes in the responsible quenching mechanism with location), will naturally lead to evolution in the spiral's morphology with time, perhaps starting from an initially spatially extended configuration and then evolving to a narrower feature around the resonance as the non-resonant component evolves away, for instance.

Maturing spirals will also tend to undergo a spectral coarsening, systematically shifting power from multi-armed, flocculent features to a global low-$m$ mode.  This is the result of the inverse dependence of the spiral lifetime on $m$; the high-$m$ components of any single perturbation naturally evolve away first, leaving an increasingly coherent structure over time.  This consolidation or inverse cascade also dictates the disk’s long-term secular appearance. Whereas a dynamically cool disk supports instabilities over a large range of scales, the heat generated by (high-$m$) waves progressively restricts the allowed range of $m$ to lower values, permanently shifting the galaxy toward a more global, coherent grand-design morphology.

The intuition gained with eq. (\ref{eq:timescale}) implies that the evolution and consolidation to low-$m$ is also an evolution ultimately towards quasi-steadiness.
As $m$ decreases, the lifetimes lengthen and the growth rates decrease (see eq.~[\ref{eq:phase1growthbeta}] or eq.~[\ref{eq:long}]).  
This stems from the intimate connection between azimuthal forcing and amplification through the donkey effect described in \S \ref{sec:exponential_caverna}.
For the lowest $m$, the growth slows enough that the potential-density phase offset remains small, preventing runaway (fast) growth from ever initiating.
This keeps $\mathcal{F}_{NR}\approx 2T_1R$ small, so that $d C_g/dR$ becomes increasingly small along the spiral, except at resonance, as is characteristic of steady spirals.
The net effect is that an effectively quasi-steady low-$m$ spiral is left behind once intermediate and high $m$ features evolve away.
In this case, we expect the quasi-steady spiral's interaction with the disk to occur over increasingly narrow resonant zones in the disk (as opposed to preceding high-$m$ spirals that once enabled changes in angular momentum and heating away from resonance).
This is precisely the regime of LBK72's adiabatic calculation.  Later in $\S$ \ref{sec:heatingradialmigration} we will examine this consolidation to quasi-steadiness in action space.  

The warming of the disk over time will lead to a similar evolution in the properties of the global spiral pattern, as high-$m$ components become increasingly disfavored (see Figure \ref{fig:pitch},  $\S$ \ref{sec:fastestwaves}).   
This overall evolution towards low-$m$, steadier spirals suggests that the LBK72 adiabatic calculation applies to a late phase of the disks' evolution.
In this light, the non-steadiness not captured by \citepalias{lynden-bell72} can be thought of as a preceding, morphologically richer stage of the spiral's evolution.
That stage can be characterized both by an increasing number of patterns and arms and as a sharpening of the arms, depending on the initial seed perturbation.
The combination of multiple $m$ to the global spiral can tighten the arms in azimuth (depending on the relative contributions of each $m$) and, once the higher $m$ components evolve away, the spiral broadens.

The transition over time from widespread non-resonant interactions between the wave and the disk to primarily resonant interactions also implies that the creation of grooves and the onset of a recurrent cycle of groove modes is a relatively late-time phenomenon.
Numerical studies of groove modes that remove all but the lowest sectorial harmonics (to isolate the groove mechanism from other sources of instabilities) are already an optmized view of that late evolution \citep[e.g.][]{sellwood19,sellwood21,sellwood22}.

\subsubsection{Comparison to Isolated N-body studies and recent simulations in the cosmological context}\label{sec:discussion_sims}
Recent studies of spiral patterns in simulated galaxies -- both within a cosmological context and in isolated, high-resolution N-body experiments -- 
show broad resemblance to the picture of fast-evolving spirals highlighted in this work.
Rather than steady and monolithic, the simulated spirals are dynamic, fast-evolving structures that are seeded through a variety of processes (including interactions and internal stimuli), depending on the host galaxy's specific environmental and internal conditions \citep[e.g.][]{donghia,quinn25,ghosh25,grand26}.
Although the individual spirals that build the global pattern evolve quickly, with lifetimes on the order of several hundred Myr, frequent triggers keep spiral activity persistent for Gyr \citep{quinn25}.

The transience of the waves, which limits the secular changes produced by any one generation, together with the anchoring of amplification to $\Omega$, will make the global spiral nexum prone to resembling a shearing pattern \citep[see also][]{sellwood14}.  But even structures that appear to be passive, non-heating material arms \citep[i.e.][]{baba13} may instead be short-lived density waves whose transience and high arm count structurally keep the disk temporarily kinematically cold.  Tools like spectrograms and longer integration windows capturing accumulated changes present revealing evidence in many cases that individually short-lived spirals evolve through the disk-spiral feedback loop, rather than simply evolving away in a shearing timescale.  
Although global, spatially extended patterns often have pattern speeds that generally decrease with radius, their presence across stellar populations with a range of ages suggest that many are indeed density-wave like in nature, rather than winding features (i.e. \citealt{quinn25}; cf. \citealt{sellwood14}).

Consistent with finite lifetimes and self-regulation, disks in the simulations of \cite{kwak25}
typically exhibit higher-$m$ (multi-armed or flocculent) structures to start but, as the disk dynamically evolves, the dominant spiral modes shift preferentially toward lower-$m$ structures, often resulting in broader, two-armed global patterns.
The addition of gas in this context has important implications for spiral rejuvenation and longevity (see also \citealt{sellwood11,sellwood14}).
As highlighted by \cite{ghosh25}, gas returns the disk to a dynamically cool state that supports higher-$m$ spirals than possible in the kinematically hotter, underlying stellar disk, it acts as a source of perturbations and it contributes cool stars to offset heating.

The growing consensus from these studies is that spirals can be excited in disks through a number of processes and often behave fundamentally like density waves.
But rather than resembling the QSSS envisioned by Lin \& Shu, any one spiral density wave is a transient feature of the disk.
A next key avenue to probe using the simulations is the non-linear evolution of individual spiral features, to test how the lifetimes are related to both the disk properties (velocity dispersion, surface density, rotation curve turnover) and spiral properties (including multiplicity, pitch angle and density contrast).
The simulations will also be valuable source of quantitative predictions for the amount secular disk changes produced by different kinds of spirals.

\subsection{On radial and vertical heating in action space}\label{sec:heatingradialmigration}
The qualities of early morphological complexity and spectral coarsening discussed in the previous section have implications for the way disk properties evolve over time.   
Spirals in dynamically cool, late-type disks (i.e. at earlier cosmic time) that include high-$m$ components  are capable of non-resonant and ILR/OLR heating. This heating gets taken over by intermediate-$m$ components as the disk warms. Eventually, the spirals in the latest disks (necessarily lower $m$ and steadier) contribute predominantly resonant heating. (Likewise, radial migration would be more widespread early on and focused at corotation later.)  
In this section we will translate these findings into action space.   We will begin by (re)examining the characteristic lifetimes most conducive to changes in radial and vertical action (or dynamical heating and vertical thickening).  Later in $\S$ \ref{sec:radazimuthal} we will consider the implications for the heating histories of galactic disks, as similarly discussed by \cite{hamilton24c}.  

\subsubsection{Resonant vs. non-resonant heating}
As an alternative to the view of heating considered in \S \ref{sec:heating} here we examine the changes in radial, azimuthal and vertical actions $\Delta J_R$, $\Delta J_\theta$ and $\Delta J_z$. 
We begin by rewriting $\Phi_1$ as a function of action $J$ and angle $\theta$ variables 
\begin{equation}
\Phi_s(\theta, J, t) =\Re{\Bigg[\sum_{n, m,l} \Phi_{n,m,l}(J, t) \, e^{i(n\theta_R + m\theta_\phi+ l\theta_z)}\Bigg]},\label{eq:phis}
\end{equation}
similar to, e.g., Hamilton (2024), in terms of the angles  $\theta_R=\theta_{R,0}+\kappa t$,$\theta_z=\theta_{z,0}+\nu t$ and $\theta_\phi=\theta_{\phi,0}+\Omega t$.  For the amplitude we adopt our Gaussian time dependence $\Phi_{n,m,l}(J, t)=\Phi_{a,n,m,l}(J) \textrm{exp}[-t^2/(2t_s^2)]$ from section $\S$ \ref{sec:saturationphase} and the $\textrm{sech}^2(z/z0)$ vertical distribution in $\S$ \ref{sec:elements} that contributes only even $l$.  

In this case, the changes in (specific) action $\Delta J_R=\int -\partial \Phi_s/\partial \theta_R dt$, $\Delta J_z=\int -\partial \Phi_s/\partial \theta_z dt$ and $\Delta J_\phi=\int -\partial \Phi_s/\partial \theta_\phi dt$ become 
\begin{equation}
\Delta J_R =\sum_{n, m,l} n \bar{\Phi}_{a,n,m,l}  t_s 
\end{equation}
\begin{equation}
\Delta J_z =\sum_{n, m,l} l \bar{\Phi}_{a,n,m,l} t_s 
\end{equation}
and
\begin{equation}
\Delta J_\phi =\sum_{n, m,l} m \bar{\Phi}_{a,n,m,l} t_s 
\end{equation}
in terms of  
\begin{equation}
\bar{\Phi}_{a,n,m,l}= \sqrt{2\pi} \Phi_{a,n,m,l} e^{-\frac{\omega_{\rm 3D}^2t_s^2}{2}} \sin(\phi_0),
\end{equation}
where
\begin{equation}
\omega_{\rm 3D} = n\kappa + m(\Omega - \Omega_p) + l\nu
\end{equation}
and
$\phi_0=n\theta_{R,0}+m\theta_{\phi,0}+l\theta_{z,0}$. 

We will take only the largest $l=0$ and $l=2$ terms, given the small contribution of higher order terms to the spiral potential's assumed vertical distribution.  In this case the lowest order terms at each $n$ and $m$ become
\begin{equation}
\Delta J_{\phi,n,m}\propto m \bar{\Phi}_{a,n,m,l=0}\sqrt{2\pi} t_s e^{-\frac{(n\kappa+m(\Omega-\Omega_p))^2t_s^2}{2}}  \label{eq:dlr}
\end{equation}
\begin{equation}
\Delta J_{R,n,m}\propto n \bar{\Phi}_{a,n,m,l=0} \sqrt{2\pi} t_s e^{-\frac{(n\kappa+m(\Omega-\Omega_p))^2t_s^2}{2}} \label{eq:dltheta}
\end{equation}
and
\begin{equation}
\Delta J_{z,n,m}\propto 2 \bar{\Phi}_{a,n,m,l=2} \sqrt{2\pi} t_s e^{-\frac{(n\kappa+m(\Omega-\Omega_p)+2\nu)^2t_s^2}{2}} \label{eq:dlz}
\end{equation}

The exponential term in all three expressions regulates the well-known link between spiral lifetime and changes in action.   The steady, long-lived spirals with large $t_s$ can produce changes in $J_R$, $J_\phi$ only at resonances, where $n\kappa+m(\Omega-\Omega_p)=0$ (see \citetalias{lynden-bell72}).  Small $t_s$, on the other hand, opens the possibility of non-resonant impulse-like changes (\citetalias{lynden-bell72}, \citetalias{van-der-wel25}).  This is precisely the regime of the intermediate and high-$m$ spirals discussed in section \ref{sec:disk_evolution} that are increasingly short-lived as $m$ increases.  Using the lifetime estimates there, we conclude that quasi-steadiness and resonant exchanges are primarily the product of low-$m$ spirals, while non-steadiness and non-resonant interactions characterize intermediate- and higher-$m$.   

The shortening of lifetimes with increasing $m$ also introduce the possibility of vertical heating.  Whereas spirals in thin disks with $\nu\gg \kappa$ are generally inefficient at producing changes in $\Delta J_z$, their impact grows as $\nu$ decreases (i.e. as the disk thickens) and $t_s$ shortens.  The detailed ratio of vertical to radial heating then depends on the vertical and radial shape of the perturbing potential and the factor $\nu t_s$.  (Spirals with a broad vertical profile induce more heating than those tightly confined to the mid-plane.) 

The link between arm number and dynamical coolness ($\S$ \ref{sec:fastestwaves}) suggests that heating is more impulse-like via short-lived spirals in cold disks (i.e. at earlier cosmic time) but focuses more towards resonances as the disk warms and the spiral spectrum becomes more steady.  Gas disks can extend the spectrum of perturbations to even higher $m$, keeping the stars exposed to impulse-like heating even after they enter a steady-spiral regime.   
Clumpy GMCs are a conventional source of vertical (and radial) heating, modeled either as the result of short lifetimes or short interaction times between orbiting stars and the perturbing GMCs.  In a forthcoming paper we propose treating gas structure as a network of filamentary high $m$ spirals, rather than as a population of GMCs, in which case the heating by gas structures can be treated  just as in the case of stellar spiral perturbations in eqs. (\ref{eq:dlr}), (\ref{eq:dltheta}) and (\ref{eq:dlz}). Characteristically high $m$ and short lifetimes (set to the dissipation timescale) in dynamically cool gas disks makes the heating non-resonant and impulse-like.  The degree to which this sort of heating is recovered in the stellar population (relative to steady, resonant heating) will depend on the  relative mass in gas and its variation over time \citep[see also][]{aumer16,blandhawthorn24,zhang25}.  

\subsubsection{The ratio of heating-to-radial migration: the importance of high-$m$ spirals}\label{sec:radazimuthal}
Disk-like morphologies and dynamics today place strong constraints on the history of secular evolution experienced in the past.
This is a point emphasized by \cite{hamilton24c}, who studied secular changes produced by spirals with different properties in numerical simulations of Mestel disks.
For many of the explored spiral properties the heating per unit radial migration is substantially larger than observed in the Milky Way \citep{frankel20}.
\cite{hamilton24c} concludes that spirals can match the low heating only if they had higher pitch angles in the past or possibly strongly tapered radial profiles, to avoid the heating at resonance (but see \citealt{dehnen25}).

To compare with the study of \cite{hamilton24c}, here we will examine the change in radial action $\Delta J_R$ relative to the change in azimuthal action $\Delta J_\theta$ specifically at the galactic mid-plane.  As in sections \ref{sec:growth_rates}-\ref{sec:disk_evolution}, we adopt the epicyclic approximation whereby $\Delta J_R=\Delta\sigma^2/\kappa$ and $J_\theta=L=RV_c$.  We start with the equivalence in eq. (\ref{eq:rate0potential}), which relates the change in velocity dispersion to the change in angular momentum produced over the spiral's lifetime, but taken without integrating over the vertical direction (as in eq. [\ref{eq:totaltorque}]).  We will also use eq. (\ref{eq:dltheta}) to write $\Delta J_\theta\approx I(t_s) t_s m\Phi_1$ where the factor $I(t_s)=\textrm{exp}[-(n\kappa+m(\Omega-\Omega_p))^2t_s^2/2]$ measures proximity to the impulse approximation.   
In this case, for an assumed $\textrm{sech}^2$ disk vertical distribution with $\rho(z=0)=\Sigma/(2z_0)$, then 
\begin{equation}
\Delta J_R\vert_{z=0}=\frac{t_s\mathcal{W}_s}{\kappa} \mathcal{F}_{\rm NR} \frac{k m}{ 4\pi  G\Sigma R z_0^{-1}} \frac{\Delta J_\theta^2\vert_{z=0}}{m^2 t_s^2 I(t_s)^2}
\end{equation}
at the mid-plane.

This simplifies to 
\begin{equation}
\left.\frac{\Delta J_R}{\Delta J_\theta}\right\rvert_{z=0}=\frac{\sqrt{2}\mathcal{W}_s}{t_sI(t_s)^2} \mathcal{F}_{\rm NR} \frac{ R z_0}{ 4\pi  G\Sigma R} \cot{i_p}\frac{\Delta J_\theta\vert_{z=0}}{L_\theta}\label{eq:jrjtheta}
\end{equation}
assuming a flat rotation curve, or to 
\begin{equation}
\left.\frac{\Delta J_R}{\Delta J_\theta}\right\rvert_{z=0}=\frac{\mathcal{F}_{NR}}{\sqrt{2}\Omega t_sI(t_s)^2}\frac{\mathcal{W}_s z_0}{ \Omega R} \cot{i_p}\frac{\Delta J_\theta\vert_{z=0}}{L_\theta}
\end{equation}
specifically in the case of a Mestel disk (for which $V_c^2=2\pi G\Sigma R$).

These expressions for the mid-plane heating $\Delta J_R/\Delta J_\theta$ show the impulse-like proportionality to $\Delta J_\theta/J_\theta$ anticipated by \cite{hamilton24c}.
They are also independent of $m$ and instead proportional to $\cot i_p$ and thus capture the dependence on pitch angle found by \cite{hamilton24c}.
With vertical integration, on the other hand, the total heating in a Mestel disk becomes
\begin{equation}
\frac{\Delta J_R}{\Delta J_\theta}=\frac{\sqrt{2}}{\Omega t_sI(t_s)^2}\frac{\mathcal{W}_s \delta_t}{ 4\Omega m} \frac{\Delta J_\theta}{L_\theta},
\end{equation}
characterized by the same $1/m$ dependence of the calculations in \S \ref{sec:quenching_heating}.

Although the precise amount of spiral heating will depend on whether it is the (vertically) integrated heating or mid-plane heating, we can conclude from \S \ref{sec:quenching_heating} or eq. (\ref{eq:jrjtheta}) that higher $m$ and lower $\cot i_p$ (higher pitch angle) will produce less heating per unit radial migration than lower-$m$, more tightly-wrapped spirals.  Although short-lived, high-$m$ spirals can heat, as $m$ increases and the azimuthal force strengthens relative to the radial force, changes in angular momentum become dominant over radial heating.  

In this light, the preferentially high-$m$ spirals in dynamically cool early disks (i.e. Figure \ref{fig:pitch}, $\S$ \ref{sec:fastestwaves}) will keep the disk relatively dynamically cool, with low heating-to-radial migration.  Eventually, the slow heating restricts $m$ and over time heating becomes enhanced.  This is the stage of accelerated compaction and heating that transforms star-forming disks into early-type galaxies \citepalias{van-der-wel25}.  

Galaxies that have survived until today with dynamical coldness in tact must be only slighlty removed from a dynamically cold past and a preference for high-$m$ spirals.  This cold evolution  presumably also reflects the influence of a dynamically cold gas component.  
Stellar disks that have reached today with a higher $\Delta J_R/\Delta J_\theta$, in contrast, are already more mature by comparison.  The stars in these systems are relatively less influenced by the gas and instead heated progressively more efficiently as high-$m$ spirals became less favored over time.

\section{Summary and Conclusions}

The origin and evolution of spiral structure in galactic disks is a longstanding puzzle with profound consequences for our understanding of how galaxies evolve over cosmic time.  In the canonical quasi-steady density wave picture \citep{lin64, lynden-bell72}, spirals are long-lived features that efficiently reshape their disk's angular momentum distribution \citep{goldreich79, binney08} through localized interactions with the disk primarily at resonances \citep{lynden-bell72}.  The contemporary version of this picture allows that the localized changes in angular momentum and heating produced by the resonant interactions \citep{sellwood02} lead to transience \citep{sridhar19,hamilton24} and stimulate new generations of waves via groove mode instability \cite{sellwood14,sellwood19}. The result is once again long-lived spiral activity, but now through a recurrent cycle of short-lived density waves \citep{sellwood91, sellwood14}. 

In this study (following up on earlier work; \citetalias{meidt24,van-der-wel25}) we widen the picture still further and explicitly incorporate non-steadiness and short lifetimes, demonstrating how non-resonant interactions emerge and contribute to transience through widespread (rather than resonant) disk secular changes.    
We use a self-consistent macroscopic hydrodynamical framework (\S\ref{sec:hydro}, \S\ref{sec:potential_perts}) that goes beyond the tight-winding limit to 
obtain a comprehensive view of collectively excited, transient spiral arms—which we term \textit{spiral nexa}—and their integrated role in disk evolution (\S\ref{sec:growth_rates}, \S\ref{sec:timedep}, \S\ref{sec:heating}, \S\ref{sec:disk_evolution}). 
By treating the stellar disk as a fluid characterized by an effective pressure, we bypass the complexities of kinetic theory \citep[e.g.,][]{fouvry15, hamilton24, hamilton25} and show how spiral properties link directly to the changes in  the disk that they bring about.  
Our macroscopic perspective captures non-adiabatic growth and decay missed by strictly secular calculations (\cite{lynden-bell72}; see  \citepalias{van-der-wel25}), bringing into view the inherently transient nature of spirals absent from traditional steady, long-lived density wave treatments \citep{lin64}, but now standard in numerical simulations \citep[e.g.][]{sellwood12,sellwood14,sellwood22,roskar12,grand13,grand26,baba13,quinn25}.  

Using a characteristic equation derived from the linearized equations of motion (\citetalias{meidt24},\citetalias{van-der-wel25}) we find (\S \ref{sec:growth_rates}) that the fundamental source of non-resonant spiral excitation is the presence of mild, large-scale gradients in a disk's mass and angular momentum distributions.  These are milder cousins of grooves, edges and tapers that give rise to strong spiral amplification \citep{zang76,toomre81,sellwood89,sellwood91,de-rijcke16} but arguably more commonplace: typical exponential galactic disks embedded in dark matter halos naturally frame mild extended angular momentum deficits, which we term \textit{cavernae}.  
Within cavernae, orbiting material responds to spiral perturbations via the ``donkey effect'' which allows the spirals to continuously accumulate stars and grow through inverse Landau damping (\citetalias{lynden-bell72, meidt24}), even away from corotation.  The result is the formation of spatially extended patterns ('spiral nexa') that transport angular momentum much more broadly than waves restricted to narrow resonance zones or grooves.   

The disk's specific large-scale gradients, which frame the architecture of the caverna, dictate the nature of the spiral nexum.  We discuss how extended, slowly rising rotation curves foster non-steady, non-resonant spirals across the disk. Conversely, steeply rising inner curves open an inner zone where inner waves are forced to remain steady.  

Crucially, the non-resonant components of spiral nexa in this framework are intrinsically self-limiting.  As they propagate, non-steady (non-adiabatic) patterns do not conserve wave action and deposit angular momentum locally into the disk (\S\ref{sec:torques}; \citetalias{van-der-wel25}).  
This non-adiabatic transport liberates orbital energy, converting it into non-circular velocity dispersion (dynamical heating).  
We derived analytical expressions that link this dynamical heating directly to observable spiral properties, demonstrating that heating efficiency scales strongly with the wave's multiplicity ($m$), pitch angle, and peak density contrast (\S\ref{sec:heating2}).  We then incorporated the heating back into our characteristic equation to depict a spiral wave's quasi-linear evolution: enhancements in velocity dispersion (and effective pressure and the Jeans length, $\lambda_J$) ultimately overwhelm the self-gravity that sustains the spiral, bringing growth to an end (\S\ref{sec:quenching_heating}).

In addition to this self-limiting nature of non-resonant heating, we identify two other quenching mechanisms that regulate the lifetimes of the resonant parts of spiral nexa.  The first is the steepening of the perturbation's own amplitude gradient (\S\ref{sec:bottomsdream}, Bottom's Dream; \citetalias{meidt24}) and the second is the local flattening of density profiles at corotation via changes in angular momentum (\S\ref{sec:quenching_corotation}; \citet{goldreich79, binney08}).  The latter is analogous to saturation via non-linear orbit trapping \citep{sellwood22, hamilton24}, which would occur alongside the other quenching mechanisms.  
Altogether, we find that these quenching mechanisms operate in tandem to limit individual spiral lifetimes to roughly 2–10 orbital periods.  

The transience of individual spiral nexa acts as the continuous engine for secular galactic evolution (\S\ref{sec:discussion}). Widespread angular momentum redistribution carves new cavernae, priming the disk for subsequent spiral generations \citep{sellwood14a}.  Those generations can be seeded by a diversity of perturbations, from molecular clouds, halo substructure, accretion events and interactions to earlier spiral episodes that trigger new perturbations through nonlinear mode coupling \citep[e.g.][]{tagger87,masset97,donghia,chiba25}.  The result is a series of recurrent spirals, in which each generation's properties is set by the evolving disk conditions produced by preceding generations.  

Over time, cumulative dynamical heating drives spectral coarsening. In young, dynamically cool systems, the short Jeans length supports high-multiplicity ($m$), open, and multi-armed spiral features, as in late-type Scd spirals.
As successive generations of these open spirals effectively heat the disk, high-$m$ perturbations are subsequently suppressed.
Over time, the structural response of the disk consolidates into lower-$m$, tighter, and increasingly coherent global patterns, as seen in early-type Sa spirals, that grow more slowly and interact with the disk over increasingly narrow resonant zones.   This evolutionary trajectory suggests that the quasi-steady, ``grand-design'' density waves described by classical theory \citepalias{lynden-bell72} are a late phase in a spiral's evolution, rather than the default state, the product of a system already matured and heated by generations of transient spiral activity.

We complete our view of spiral evolution (\S\ref{sec:heatingradialmigration}) by examining the ratio of radial dynamical heating to radial migration ($\Delta J_R / \Delta J_\theta$).  The high-$m$, open spirals characteristic of cold early disks produce significantly less heating per unit radial migration than late-stage, tightly wound features \citep{hamilton25}.
This implies that the observed kinematically cold state of the Milky Way's thin disk \citep{frankel20} may be a natural consequence of an evolutionary history dominated by high-$m$ spiral nexa operating within a gas-rich environment.  Dynamically warm, compact early-type galaxies, on the other hand,  would represent a more mature state reached once spectral coarsening has left behind intermediate- and low-$m$ spiral features \citepalias{van-der-wel25}.

The results of this study suggest that spiral arms are not merely passive indicators of a host galaxy's underlying potential. 
They are active, self-regulating agents that systematically drive exponential galactic disks toward their maximum entropy configurations through a continuous cycle of growth, angular momentum transport, dynamical heating, and quenching.

\end{document}